\documentclass[conference]{IEEEtran}
\IEEEoverridecommandlockouts
\usepackage{cite}
\usepackage{amsmath,amssymb,amsfonts, bm}
\usepackage{algorithmic}
\usepackage{graphicx}
\usepackage{textcomp}
\usepackage{xcolor}
\usepackage{stfloats}
\usepackage{algorithm}
\usepackage{diagbox}
\usepackage{makecell}
\usepackage{setspace}
\usepackage{float}
\usepackage{caption}
\usepackage{booktabs} 
\usepackage{enumerate}

\def\BibTeX{{\rm B\kern-.05em{\sc i\kern-.025em b}\kern-.08em
    T\kern-.1667em\lower.7ex\hbox{E}\kern-.125emX}}
\begin{document}

\title{Joint Channel Assignment and Power Allocation for Multi-UAV Communication
}

\author{\IEEEauthorblockN{
			Lingyun Zhou, 
		Xihan Chen, Mingyi Hong, Shi Jin and Qingjiang Shi}
	
	\thanks{
		L. Zhou and Q. Shi are with the School of Software Engineering, Tongji University, Shanghai 200092, China (e-mail: 1911561@tongji.edu.cn;  qing.j.shi@gmail.com). 		
	
	X. Chen is with the Department of Information Science and Electronic Engineering, Zhejiang University, Hangzhou 310000, China (e-mail: chenxihan@zju.edu.cn).
	
	M. Hong is with the Department of Electrical and Computer Engineering, University of Minnesota, Minneapolis, MN 55455, USA (e-mail: mhong@umn.edu).		
	
	S. Jin is with the National Mobile Communications Research Laboratory, Southeast University, Nanjing 210096, China (e-mail: jinshi@seu.edu.cn). 
	
	An early version of this paper has been published in IEEE 21th International Workshop on Signal Processing Advances in Wireless Communications (SPAWC 2020) \cite{ZLY}.	
	}
}

\maketitle

\begin{abstract}
Unmanned aerial vehicle (UAV) swarm has emerged as a promising novel paradigm to achieve better coverage and higher capacity for future wireless network by exploiting the more favorable line-of-sight (LoS) propagation. To reap the potential gains of UAV swarm, the remote control signal sent by ground control unit (GCU) is essential, whereas the control signal quality are susceptible in practice due to the effect of the adjacent channel interference (ACI) and the external interference (EI) from radiation sources distributed across the region. To tackle these challenges, this paper considers priority-aware resource coordination in a multi-UAV communication system, where multiple UAVs are controlled by a GCU to perform certain tasks with a pre-defined trajectory. Specifically, we maximize the minimum signal-to-interference-plus-noise ratio (SINR) among all the UAVs by jointly optimizing channel assignment and power allocation strategy under stringent resource availability constraints. According to the intensity of ACI, we consider the corresponding problem in two scenarios, i.e., Null-ACI and ACI systems. By virtue of the particular problem structure in Null-ACI case, we first recast the formulation into an equivalent yet more tractable form and obtain the global optimal solution via Hungarian algorithm. For general ACI systems, we develop an efficient iterative algorithm for its solution based on the smooth approximation and alternating optimization methods. 
Extensive simulation results demonstrate that the proposed algorithms can significantly enhance the minimum SINR among all the UAVs %as compared to the existing solutions 
and adapt the allocation of communication resources to diverse mission priority.

\end{abstract}


\section{INTRODUCTION}
For the fifth-generation (5G) communication networks, there is a urgent need to improve the system performance limit to accommodate the ever increasing data traffic, which poses very stringent requirements on both the radio resources and the existing communication  infrastructures \cite{5G}. 
With benefits such as low cost, high maneuverability, and on-demand deployment, unmanned aerial vehicle (UAV) enabled communication serves is very promising in achieving better coverage and higher capacity for future wireless network \cite{UAVinto1,UAVinto2,UAVinto3,UAVinto4}.
%With benefits such as low cost, high maneuverability, and on-demand deployment, Unmanned Aerial Vehicles (UAVs) have been widely used in wireless communications system\cite{UAVinto1}. 
In particular, UAV autonomous surveillance is regarded as the most promising application in both civil and military fields, where a group of UAVs cooperate with each other and perform inspection tasks in a specific geographical area. 
The effective collaboration among the UAVs in a swarm not only makes up for the limited hardware capability of a single UAV, but also improves the fault tolerance of the whole system, thereby making it possible to complete missions in a cost-effective manner. 
%As compared to the single-UAV technology, swarm technology allows a number of UAVs to be interconnected and controlled as a whole, with the ability to collaboratively perform missions and efficiently react to dynamic environments. 
%The cooperative operation of UAV swarm makes up for the limited capability of single UAV, improves the fault tolerance of the system, and makes tasks more efficient. 
Due to these desirable features, the UAV swarm technology have recently drawn considerable interests from the both academic and industrial communities \cite{UAVinto5,UAVinto6,UAVinto7}.
%To enable swarm technology, some relevant wireless communication issues should be first addressed, which have received a great deal of attentions in recent years\cite{UAVinto2,UAVinto3,UAVinto4}.    

Depending on the role of UAVs in the entire network, two different lines of research can be identified in the literature, namely UAV-assisted communication and cellular-enabled UAV communication. In UAV-assisted communication, UAVs serve as new aerial communication platforms to provide  services for the terrestrial.
To alleviate the performance bottleneck caused by the ``doubly near-far'' phenomenon, the authors of \cite{relate_xie} invoked wireless power transfer (WPT) techniques in the multi-UAV network, and considered the joint optimization of trajectory and resource allocation to maximize the system throughput, subject to both UAVs' speed and energy neutrality constraints. 
To provide remote terminals with connectivity opportunities for hazard detection and disaster recovery, the authors in \cite{relate_Elmagid} advocated the usage of a novel UAV relay scheme to enhance the coverage and capacity of terrestrial wireless networks.
Furthermore, two emerging Internet of Things (IoT) services, i.e., data collection and information dissemination, were respectively enabled in the UAV network \cite{relate_zeng, relate_lyu}. Specifically, the authors in \cite{relate_zeng} investigated an optimal UAV trajectory to minimize the energy consumption while collecting all the desired data from the ground users. To enable a set of ground terminals to share the allotted spectrum in the most effective fashion, the authors in \cite{relate_lyu} conceived a novel cyclical multiple access (CMA) scheme according to the variations of UAV trajectory, and investigated the fundamental tradeoff between the dissemination delay and the system throughput. 
Inspired by the need to accommodate latency-sensitive and computation-intensive emerging applications, the authors in \cite{relate_MEC} combined the mobile edge computation (MEC) technique with the UAV network, and subsequently devised a powerful resource allocation strategy to maximize the system energy efficiency.

%showed an analytical approach on the energy efficiency maximization problem for a multi-UAV communication system where each UAV is deployed to collect data from a group of ground users,  

%afixed-wingUAVisdeployedtocollectinformation from a group of distributed ground terminals

%the deployment problem of static-UAVs with modeling geographical location as a three-dimensional coordinate system, and considered to maximize network throughput while satisfying the cellular network resource constraints. 
%Two practical studies for three-dimensional coordinate placement were proposed in \cite{relate2} and \cite{relate3}, namely, altitude and horizontal optimization.
%The work in \cite{relate2} showed an analytical approach on the optimal altitude of UAV BSs for providing maximum coverage for service users under the constraint of maximum transmission power. Meanwhile, \cite{relate3} presented a horizontal positions optimization scheme to minimize the number of required BS to provide services for a given set of ground terminals.  
%Besides the study of UAV-assisted cellular communication, devising the high-performance communications between the BGS and UAVs can effectively solve the challenges from explosive growth of the number of UAVs and the demand for new applications. 
%Over other hand, cellular-enabled UAV communication system, multi-UAV can be perfectly operated by GBSs to perform their own missions (e.g.,cargodelivery, video surveillance). 

On the other hand, multi-UAV can be perfectly manipulated by ground control units (GCUs) to perform their own missions (e.g.,cargodelivery, video surveillance) in cellular-enabled UAV communication system.
In such a scenario, the control signal reception of UAV is not only affected by the quality of communication links, but also susceptible to any potential interferences. 
Hence, fully exploring the resource allocation and interference mitigation in cellular-enabled UAV network can provide effective ways to improve the communication performance. 
In particular, the authors in \cite{relate_xue} proposed a joint time-frequency scheduling and power allocation design to guarantee the reliable signals reception in the uplink transmission, where a number of UAVs are controlled by a GCU to carry out missions. 
To minimize the task execution time while ensuring the accurate information retrieval with high probability, the authors of \cite{relate_zhang} investigated the UAV trajectory optimization design, subject to a minimum received SNR constraint, the UAV’s initial and final location constraints, as well as the maximum speed constraint. 
Furthermore, the authors in \cite{relate_mei} considered the joint optimization of UAV-cell association and transmit power control to maximize the network throughput, in the presence of multicell interference caused by the increased line-of-sight (LoS) air-to-ground channels.
Meanwhile, the authors in \cite{relate_liu} proposed a novel interference cancellation strategy for the uplink multi-antenna UAV communication system, where the limited backhaul links among adjacent GCUs are fully utilized to eliminate the UAV’s uplink co-channel interference and further maximize the network throughput.
\begin{figure}[!t]
	\centering
	\includegraphics[width=8.9cm,height=2.9in]{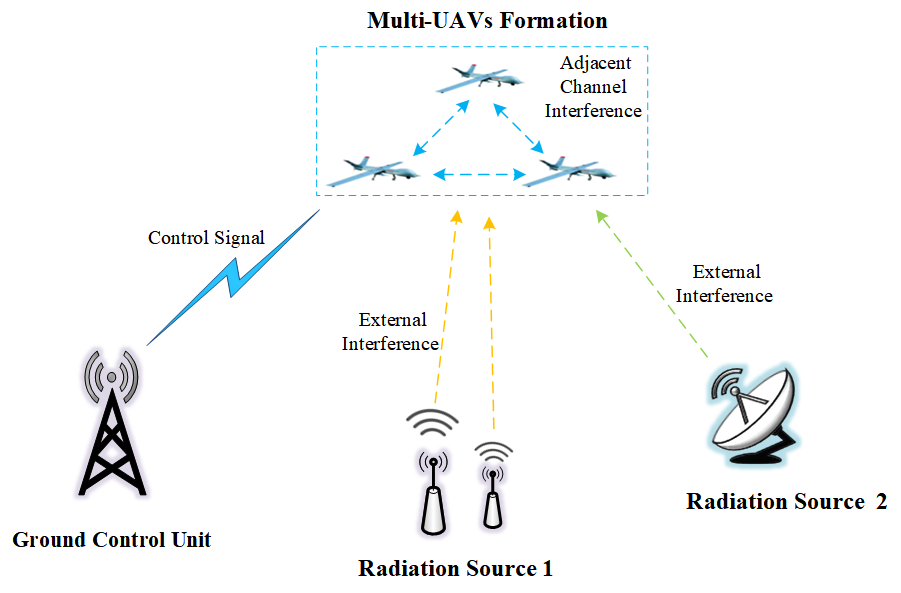}
	\caption{A multi-UAV communication scenario. The quality of wireless link is affected by ACI and EI.} \label{fig:UAV_system}
\end{figure}

%To complete missions efficiently, UAVs in a swarm need to interact with other aerial nodes quickly and safely, especially in highly dynamic dense wireless network environments. 
However, to the best of our knowledge, some important technical challenges have not been well investigated in the existing literature.
First, UAVs should correctly receive the control signals and quickly follow the instructions from the GCU, which further requires a reliable low-latency air interface and vast communication resources. 
However, in practice the available network resources are usually limited, which further aggravates the potential interference in the communication system. In fact, the characteristics of various potential interference for different missions can be quite different. As such, it is imperative to devise an efficient resource allocation strategy based on the distinctive characteristics of different interference, thereby enabling more efficient and reliable data transmission. 
Finally, most missions can be divided into several different sub-tasks according to the diverse functionality requirements, and different sub-tasks allocated to each UAV have various priorities. Hence, it is essential to take the priorities of different sub-tasks into account when conceiving the resource coordination strategies in the multi-UAV network.
Motivated by the above concerns, this paper conceives priority-aware  resource allocation strategy for the efficient control and reliable communication in a multi-UAV communication network, where a GCU controls multiple UAVs through a \emph{limited number of frequency bands} to perform some tasks with a pre-defined trajectory.
%Motivated by the above concerns, this paper considers priority-based multi-UAV resource coordination for efficient control and reliable communication in a cellular-enabled network. Specifically, we study a scenario in which a GBS controls multiple UAVs through a limited number of frequency bands to perform some tasks with a pre-defined trajectory. 
As illustrated in Fig. \ref{fig:UAV_system}, the quality of wireless communication link between GCU and each UAV is not only affected by adjacent channel interference (ACI) but also by external interference (EI) from radiation sources.
%As shown in Fig. \ref{fig:UAV_system}, in this scenario, the quality of wireless link between GBS and each UAV is not only affected by adjacent channel interference (ACI) but also by external interference (EI) from radiation sources. 
The goal of this paper is to jointly design channel assignment and power control to reduce the impact of interference and improve the quality of control signals. It is quite challenging to globally solve the resultant optimization problem, due to the highly non-convex and non-smooth nature of objective function, as well as the intricately coupled constraints. The main contributions of this paper are summarized as follows:

\begin{itemize}
	\item We present the basic model for a multi-UAV communication system. A max-min-fairness problem is then formulated to improve the quality of the received signals by joint channel assignment and power allocation under stringent resource availability constraints. According to the intensity of ACI, we consider the corresponding problem in two scenarioes, i.e., Null-ACI and ACI communication systems.	
	%	We derive a max-min-fairness problem for improving the quality of the received control signals by jointly considering the channel assignment and power allocation under stringent resource availability constraints. This formulation can effectively combat interference and improve the quality of service in multi-UAV communication system. However, this problem involves mixed continuous variables (power allocation) and discrete variables (channel selection), which belongs to  NP-Hard problem. 
	\item By exploiting the special structure of Null-ACI case, we first recast this non-convex and non-smooth optimization problem into an equivalent but more tractable form. We then propose a joint optimization algorithm to obtain its global solution based on Hungarian method \cite{Hungarian}. 
	\item For general ACI system, we first transform the original problem into a compact %yet more enunciable matrix form, 
	form, and develop an efficient iterative algorithm for its solution based on the smooth approximation and alternating optimization methods.
	%	\item According to the strength of ACI, we mainly consider the corresponding problem in two scenarioes: Null-ACI communication system and ACI communication system. For the case of Null-ACI communication system, we simplify the max-min-fairness problem to an equivalent minimization problem. Then we transform the formulation into a minimization channel assignment problem and adopt Hungarian algorithm to optimize it. For the case of ACI communication system, we propose an equivalent reformulation of the original problem by expressing the challenging model into a tractable matrix form. Next, we develop an efficient iterative algorithm based on smoothing and alternating optimization to solve this mixed-integer problem by utilizing the gradient projection and eigenvalue decomposition techniques.
	\item We perform extensive experiments under various parameter configurations. Numerical results clearly show the efficacy of the proposed algorithm, whilst providing some fundamental understanding and design guidelines for multi-UAV communication system.
	%	validate that the proposed algorithms can significantly enhance network performance as compared to the existing solutions and adapt the allocation of communication resources to diverse mission priority.
	%	We perform extensive experiments under various parameter configurations. The simulations clearly show the efficacy of the proposed algorithm. 
	%	The results also shed light on the design and performance analysis of the multi-UAV communication system.
\end{itemize}

The reminder of this paper is organized as follows. Section II describes the  multi-UAV communication model and formulates the resource allocation problem of interest. Section III considers a special scenario and proposes an efficient algorithm to obtain the global optimal solution. In Section IV, a low-complexity iterative joint channel assignment and power allocation algorithm is proposed for reliable communication with ACI system. The simulation results are provided in V. Finally, this article is concluded in Section VI.

\emph{Notations:} Throughout this paper, scalars are denoted by lower case or italic letters, vectors are denoted by boldface lower case, and matrices are denoted by boldface upper case letters. The space of $M\times N$ real matrices is expressed as $\mathbb{R}^{M\times N}$. For a matrix $\mathbf{A}$, $\mathbf{A}^{T}$, $\mathbf{A}^{-1}$ and $\lambda_{\max}(\mathbf{A})$ denote its transpose, inverse, and the maximum eigenvalue, respectively.

%Given the UAV’s trajectory coordinates, flight velocity vectors and interference information, the central processor of the control BS could calculate and deduce by UAVs’ current working state, then the resource allocation results are planed in advance. 
\section{SYSTEM MODEL}
\subsection{Network Architecture and Channel Model}
\begin{figure}[!t]
	%\centerline{\includegraphics{UAV_Scenario.png}}
	\centering
	\includegraphics[width=9cm,height=2.6in]{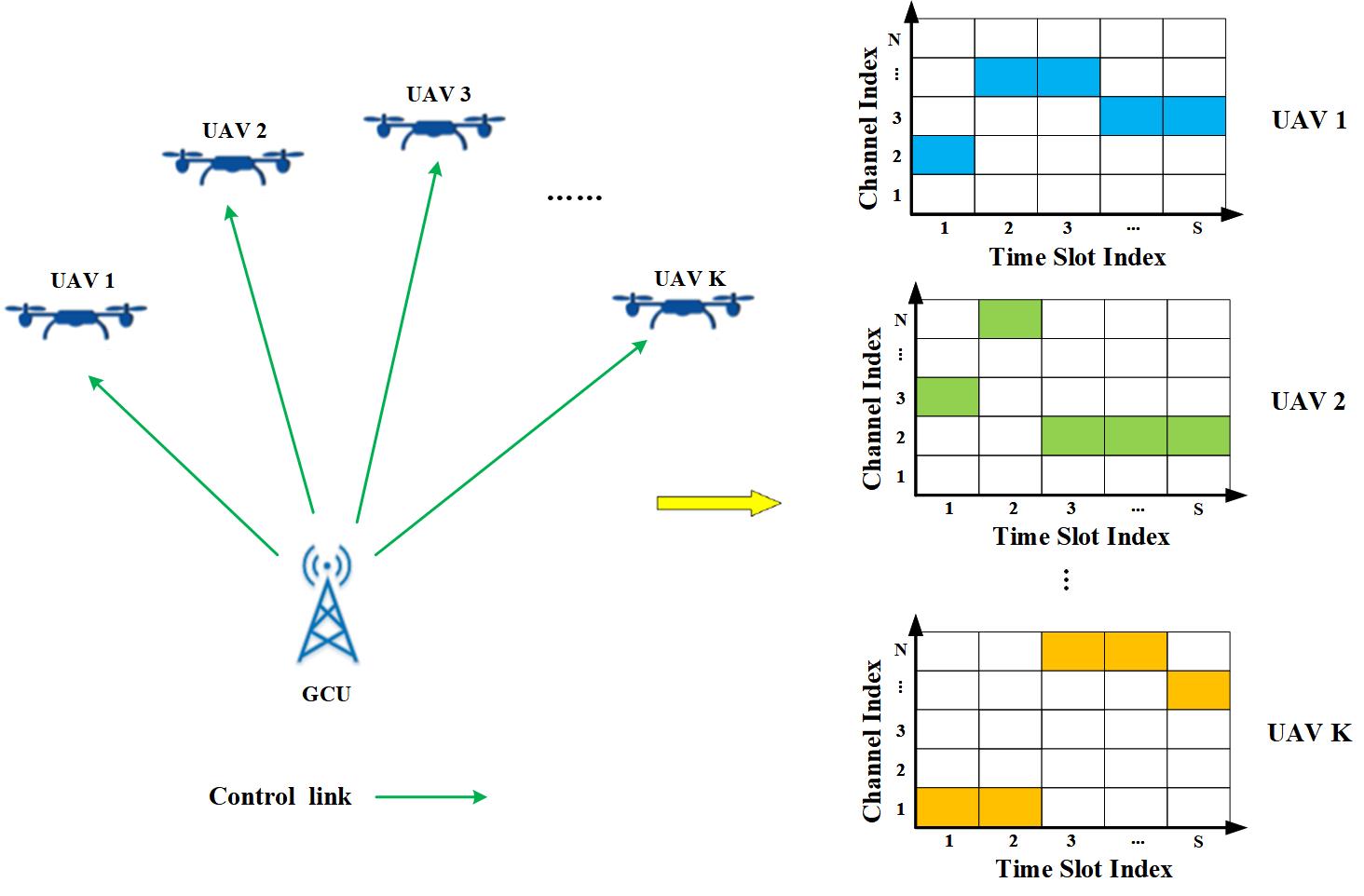}
	\caption{A GCU controls multiple UAVs to perform some tasks with a pre-defined trajectory. Each UAV is assigned with only one channel, and different UAVs access different channels.} \label{fig:UAV_Communication}
\end{figure}

As shown in Fig. \ref{fig:UAV_Communication}, we consider an uplink wireless UAV communication scenario where a GCU controls multiple UAVs to perform certain task with a pre-defined trajectory. In such a scenario, the GCU first assigns channels to UAVs from a set of limited frequency bands and then sends control signals to multiple UAVs at each time slot. 
Meanwhile, UAVs periodically feed back their information so that the GCU can not only acquire three-dimensional (3D) flight coordinates, but also perceive channel state information (CSI) of UAV swarm. 
We use $\mathcal{K} \triangleq\{1, \ldots, K\}$ to denote the set of UAVs, $\mathcal{N} \triangleq\{1, \ldots, N\}$ the set of communication channels, and $\mathcal{S} \triangleq\{1, \ldots, S\}$ the set of time slots. We assume that the length of time slot is chosen to be sufficiently small such that the UAV's location remains unchanged within every time slot. 

%In general, the existence of ACI is mainly attributed to the following two reasons: First, due to the limited performance of UAV receiver filter, the bit error rate (BER) of the communication system increases. Second, the side lobe effect of the GBS antenna may bring mutual interference between the adjacent channels.

In this paper, we consider two kinds of interference sources to distinguish different transmission scenarios. 
First, multiple UAVs are likely to occupy adjacent channels simultaneously due to the shortage of spectrum resources in practice. Consequently, owing to the limited performance of receiver filters, it inevitably results in the ACI, which may cause undesired influence on the reliable communication between UAVs and the GCU \cite{ACI}. For example, a GCU simultaneously sends control signals to UAV $k$ and $m$ via adjacent channels, $d_{s, k}$ and $d_{s, m}$ are the distance between the GCU and two UAVs at time slot $s$, respectively. If $d_{s, k} > d_{s, m}$, the high power signals for UAV $k$ would leak into adjacent channels and interfere with the communication between the GCU and UAV $m$. 
Second, we consider that the EI also possibly exists due to the radiation sources (e.g., eavesdroppers, enemy radar, and many others.) distributed across the region (in which the planed UAVs' trajectory is located) and its intensity highly depends on the number, type, and location. In such a scenario, it is important to devise efficient resource allocation strategies based on the distinctive characteristics of different interference sources, thereby enabling more efficient and reliable data transmission for carrying out the task.

%Here, two kinds of interference sources should be considered.
%On one hand, due to the shortage of spectrum resources, it is likely that multiple UAVs occupy adjacent channels simultaneously, so the adjacent channel interference (ACI) may cause undesired influence on the system \cite{ACI}. For example, a GBS simultaneously sends control signals to UAV $u_1$ and $u_2$ via adjacent channels, $d_{u_1}$ and $d_{u_2}$ are the distance between the GBS and two UAVs, respectively. If $d_{u_1} > d_{u_2}$, the high power signals for UAV $u_1$ would leak into adjacent channels and interfere with the communication between the GBS and UAV $u_2$. 
%On the other hand, we consider that the external interference (EI) also possibly exists due to radiation sources distributed across the region (in which the planed UAVs' trajectory is located). Specifically, severe interference may be generated to the GBS-to-UAV communication links from the external radiation sources (e.g., eavesdroppers, enemy radar) that are transmitting at the same channel as the GBS. The aggregate EI mainly depends on the number, location and intensity of the radiation sources.
%In the above scenario, it is meaningful to design reasonable resource coordination strategy to combat the interference and improve the quality of control signals.

In the sequel, we elaborate the ACI model. First, we introduce $\mu_{{f_{1}}{f_{2}}}$ to characterize the interference correlation between the two spectrums specified by $f_{1}$ and $f_{2}$, which strictly satisfies the following properties:
\begin{equation}\label{eq:ACI}
\begin{cases}
0 \leq \mu_{f_{1}, f_{2}} \leq 1,\quad \mu_{f_{1}, f_{2}}=\mu_{f_{2}, f_{1}},\\
%\mu_{f_{1}, f_{2}}=\mu_{f_{2}, f_{1}}, \\
\ \mu_{f_{1}, f_{2}} = 1, \quad \text{if} \quad |f_{1} - f_{2}| = 0, \\
\ \mu_{f_{1}, f_{2}} \rightarrow 0, \quad \text{if} \quad |f_{1} - f_{2}|\rightarrow \infty, \\
\end{cases}
\end{equation}
where $\mu_{f_{1}, f_{2}}=\mu_{f_{2}, f_{1}}$ indicates the symmetric property of interference correlation coefficients,
$|f_{1} - f_{2}| = 0$ implies that two UAVs simultaneously occupy the same channel, and $|f_{1} - f_{2}|\rightarrow \infty$ shows that the channel $f_{1}$ and $f_{2}$ are sufficiently separated. 
Note that the interference correlation coefficient $\mu_{f_1f_2}$ is proportional to the intensity of ACI and can be measured in practical systems \cite{relate_xue}.
%resource coordination for efficient control and reliable communication in a BS-controlled-UAV network
%which renders the interference management as a challenging problem to solve.
%Therefore, it is meaningful to design reasonable resource allocation scheme to combat the interference and improve the quality of control signals.

%We use $\mathcal{K} \triangleq\{1, \ldots, K\}$ to denote the set of UAVs, $\mathcal{N} \triangleq\{1, \ldots, N\}$ the set of communication channels, and $\mathcal{S} \triangleq\{1, \ldots, S\}$ the set of time slots. We assume that the length of time slot is set to be sufficiently small such that the UAV's location remains unchanged within every time slot. To fully utilize the scarce spectrum resources, the available bandwidth are usually tightly segmented, thus the aggregate ACI could not be ignored when multi-UAVs access spectrum of adjacent channels simultaneously. In general, the existence of ACI is mainly attributed to the following two reasons: First, due to the limited performance of UAV receiver filter, the bit error rate (BER) of the communication system increases. Second, the side lobe effect of the GBS antenna may bring mutual interference between the adjacent channels.
%To model the ACI, we use $\mu_{{f_{1}}{f_{2}}}$ to express the interference coefficient between the two spectrums specified by $f_{1}$ and $f_{2}$. Theoretically, it has the following features
\newcounter{TempEqCnt1} % 创建临时变量TempEqCnt
\setcounter{TempEqCnt1}{\value{equation}} % 将当前公式序号 赋给TempEqCnt
\setcounter{equation}{6} % 当前公式序号变为x，x等于长公式应有的序号减1.
\begin{figure*}[b]
	\hrulefill % 在公式之前加上一条分割线，长度为页面宽度
	\begin{equation}\label{eq:SINR1_model}
		\gamma_{s, k} (\mathbf{A}, \mathbf{P}) = \frac{ \sum\limits_{n=1}^{N} a_{s, k}^{n} p_{s,k} c_{s,k}f_n^{-2}}{\sum\limits_{n=1}^{N} \sum\limits_{j=1}^{N}\sum\limits_{m\not=k}^{K}  a_{s, k}^{n} a_{s, m}^{j}\mu_{f_{n} f_{j}} p_{s,m}  c_{s,k}f_{j}^{-2}+ \sum\limits_{n=1}^{N} a_{s, k}^{n} \sigma_{s,k,f_n}^2}
	\end{equation}
\end{figure*}
\setcounter{equation}{\value{TempEqCnt1}} % 把TempEqCnt中存的公式序号赋回给当前
Since obstacles surrounding the GCU may have irregular shapes and random locations, the wireless channel from the GCU to the UAV are dominated by either LoS or non-line-of-sight (NLoS) propagation \cite{LoS1, LoS2}. Specifically, the path loss exponent of NLoS link $\eta_{L}$ is usually higher than that of LoS link $\eta_{N}$ due to shadow effects and the obstacle penetration, corresponding to the less favorable propagation in the ground-to-air communication. Accordingly, the path loss between the GCU and UAV $k$ for the LoS and NLoS at time slot $s$ are given by \cite{ZLY}
\begin{equation}
\mathrm{P L}_{s, k}(\eta)= 20 \lg \left(f_{s,k}\right)+20 \lg \left(d_{s,k}\right)+32.4 + \eta 
\end{equation}

\noindent with $\eta \in \left\{ \eta_{L}, \eta_{N}\right\}$,  $f_{s, k }\in\left\{ f_{1}, f_2,\ldots,  f_{N}\right\} $ represents the channel assigned to UAV $ k $ at time slot $s$ and $d_{s,k} $ denotes the distance between the UAV swarm and GCU. Consequently, the channel gain between the GCU and UAV $ k $ on the channel $f_{s,k}$ at time slot $s$ can be expressed as
\begin{equation}\label{eq:channel_gain}
g_{s, k}(\eta)=\frac{1} {\mathrm{P L}_{s, k}(\eta)}=c_{s,k}(\eta)f_{s, k }^{-2}
\end{equation}

\noindent with $c_{s,k}(\eta)\triangleq10^{-32.4-\eta} \times d_{s,k}^{-2}$.  

The uplink transmission power of the GCU for UAV $k$ at time slot $s$ is denoted by $p_{s,k}$, which is subject to the power budget constraint $\sum_{k=1}^{K} p_{s,k} \leq P_{\max }$, $\forall s$ with $P_{\max }$ being the maximum available transmission power of the GCU. For the sake of notational simplicity, we introduce a binary variable $a_{s, k}^{n}$ serving as the scheduling indicator for channel $n$ and UAV $k$ at time slot $s$, i.e., $a_{s, k}^{n} = 1$ if channel $n$ is assigned to UAV $k$ at time slot $s$ and $a_{s, k}^{n} = 0$ otherwise. Note that since the available frequency spectrum is usually limited in practice, we assume that in each time slot, each UAV only occupies one channel, while each channel is only assigned to at most one UAV. Such a requirement yields the following constraints:
\begin{equation}\label{eq:constrain1}
\sum\limits_{k=1}^{K} a_{s, k}^{n} \leq 1,\quad \forall n\in \mathcal{N},\quad \forall s \in \mathcal{S}, 
\end{equation}
\begin{equation}\label{eq:constrain2}
\sum\limits_{n=1}^{N} a_{s, k}^{n} = 1,\quad \forall k\in \mathcal{K},\quad \forall s \in \mathcal{S}, 
\end{equation}
\begin{equation}\label{eq:constrain3}
a_{s, k}^{n} \in\{0,1\}, \quad \forall k \in \mathcal{K}, \forall n \in \mathcal{N},s \in \mathcal{S}.
\end{equation}
Based on the above discussion, we can respectively rewrite $f_{s, k}^{-2}$ in \eqref{eq:channel_gain} and $\mu_{f_{s, k} f_{s, m}}$ in \eqref{eq:ACI} as $\sum\nolimits_{n=1}^{N} a_{s, k}^{n} f_n^{-2}$ and $\sum\nolimits_{n=1}^{N} \sum\nolimits_{j=1}^{N} a_{s, k}^{n} a_{s, m}^{j}\mu_{f_{n} f_{j}}$ for ease of exposition. Let $\mathbf{A} \triangleq \left\{a_{s, k}^{n} | s \in \mathcal{S}, k \in \mathcal{K}, n \in \mathcal{N} \right\}\in \mathbb{R} ^{S \times K\times N} $ denote the channel assignment matrix at the GCU over the whole flight duration and $\mathbf{P}\triangleq\left\{p_{s,k} | s \in \mathcal{S},k \in \mathcal{K}\right\}  \in \mathbb{R}^{S \times K}$ denote the aggregate power allocation matrix. Using the above notations, the received signal-to-interference-plus-noise ratio (SINR) at UAV $k$ in the presence of ACI and EI  at time slot $s$ is defined in \eqref{eq:SINR1_model} as displayed at the bottom of this page, where $\sigma_{s,k, f_{n}}^2$ is the power of the EI plus the additional white Gaussian noise (AWGN) on the channel $f_{n}$ at UAV $k$ in time slot $s$. The first term in the denominator of \eqref{eq:SINR1_model} stands for the ACI caused by the transmissions of all the other UAVs at time slot s.
Hence, the average received SINR of UAV $k$ over the $S$ time slots can be expressed $\gamma_{k} (\mathbf{A}, \mathbf{P}) = \frac{1}{S}\sum_{s=1}^S \gamma_{s,k} (\mathbf{A}, \mathbf{P})$.

\subsection{Problem Formulation}
It is readily seen that, the higher the SINR is, the better the communication quality of the UAV network can be achieved. Hence, it is expected to keep the SINR levels of all UAVs as high as possible. Motivated by this, we adopt the max-min-fairness objective in order to improve the worst average SINR among all UAVs by jointly optimizing the channel assignment (i.e., $\mathbf{A}$) and power allocation (i.e., $\mathbf{P}$) over the whole flight duration, under some practical constraints. Moreover, we introduce a weight factor $\alpha_{k}$ for prioritizing UAV $k$’s mission. Considering all time slots, the overall problem formulation considered in this paper can be mathematically formulated as
\setcounter{equation}{7}  
\begin{subequations}\label{eq:SINR2_model}
	\begin{align}
	\max_{\mathbf{A}, \mathbf{P}} \quad &\min_k \alpha_{k}\gamma_{k} (\mathbf{A}, \mathbf{P}) &\\
	\mbox{s.t.}\quad 
	&\sum\limits_{k=1}^{K} a_{s, k}^{n} \leq 1,\quad \forall n\in \mathcal{N},\quad \forall s \in \mathcal{S},     \label{subeq:SINR2_model_b}\\
	&\sum\limits_{n=1}^{N} a_{s, k}^{n} = 1,\quad \forall k\in \mathcal{K},\quad \forall s \in \mathcal{S},      \label{subeq:SINR2_model_c} \\
	&a_{s, k}^{n} \in\{0,1\}, \quad \forall k \in \mathcal{K}, \forall n \in \mathcal{N},s \in \mathcal{S}  \label{subeq:SINR2_model_d}\\
	&\sum\limits_{k=1}^{K} p_{s, k} \leq P_{\max },\quad \forall s \in \mathcal{S}, \label{subeq:SINR2_model_e}\\
	&0 \leq p_{s, k}, \quad \forall k \in \mathcal{K}, \quad \forall s \in \mathcal{S}.  \label{subeq:SINR2_model_f}
	\end{align}
\end{subequations}

Note that there are several challenges in solving problem \eqref{eq:SINR2_model} optimally, elaborated as follows. First,  the channel assignment matrix $\mathbf{A}$ and the power allocation matrix $\mathbf{P}$ are intricately coupled in the objective function, due to the presence of ACI. Second, the max-min fairness utility renders problem non-differentiable and non-convex. Moreover, the channel assignment indicator $a^n_{s,k}$ is a discrete binary variable, which makes the feasible-set non-convex. In short, we are faced with a mixed-integer nonlinear programming (MINP) problem, which is usually considered as NP-hard. In the next two sections, we propose two efficient algorithms to solve problem \eqref{eq:SINR2_model} in the Null-ACI and ACI cases, respectively. 
 
%It is observed that problem \eqref{eq:SINR2_model} is a mixed-integer non-convex problem and thus difficult to be solved optimally. The main challenge lies in the non-trivial relationship between the interference pattern and uplink sheduling strategy. Specifically, due to the presence of ACI, a frequency decision for a pair of UAVs are coupled with each other. For example, a particular frequency assignment $f_{s,k}$ has strongly influence on the other frequency assignment $f_{s,m}$. Note that in practice, the internal interference is may be much smaller than the external interference, e.g., a few watts versus dozens of watts, and in many cases it could be ignored. To draw essential insights into our proposed model, according to the strength of ACI, we mainly consider the corresponding problem in two scenarioes: reliable communication for NO-ACI system and reliable communication for ACI system. 

\section{Efficient Resource Allocation for Null-ACI System}
In this section, we assume that there is no ACI in the considered multi-UAV system. This assumption is reasonable when the hardware capabilities of UAVs' transceiver are strong or the interference source is dominated by the EI so that the ACI can be neglected without significant performance loss as in many existing works \cite{relate_xie, relate_Elmagid, relate_zeng}. By stipulating this assumption, we here focus on considering the efficient resource allocation for Null-ACI system. 
Such problem is also meaningful in the sense that it can give an upper performance bound for the which will be elaborated in Section IV.

In the absence of the ACI, it immediately follows from \eqref{eq:SINR1_model} that the received SINR of UAV $k$ at time slot $s$ can be expressed as
\begin{equation}\label{eq:SINR1_nullACI}
\gamma_{s, k}^{\mathrm{null}} (\mathbf{A}, \mathbf{P}) = \frac{ \sum\limits_{n=1}^{N} a_{s, k}^{n} p_{s,k} c_{s,k}f_n^{-2}}{\sum\limits_{n=1}^{N} a_{s, k}^{n} \sigma_{s,k,f_n}^2}.
\end{equation}
Therefore, the resulting joint channel assignment and power allocation problem is given by
\begin{subequations}\label{eq:model_nullACI}
	\begin{align}
	\max_{\mathbf{A}, \mathbf{P}} \quad &\min_k \frac{1}{S}\sum_{s=1}^S \alpha_{k} \gamma_{s, k}^{\mathrm{null}} (\mathbf{A}, \mathbf{P}) \label{eq: model_nullACI_a} &\\
	\mbox{s.t.}\quad 
	& \text{\eqref{subeq:SINR2_model_b}, \eqref{subeq:SINR2_model_c}, \eqref{subeq:SINR2_model_d}, \eqref{subeq:SINR2_model_e}, \eqref{subeq:SINR2_model_f}}.
	\end{align}
\end{subequations}

Note that problem \eqref{eq:model_nullACI} cannot be directly solved due to the accumulation of the received SINR of each UAV across the different time slots and the resulting multiple fractional coupling. 
However, it can be observed that prioritized received SINR of each UAV at the specific time slot is independent of that in different time slot, which allows decomposing the complicated overall problem \eqref{eq:model_nullACI} into a series of parallel subproblems across different time slot. As a result, we only need to focus on the joint optimization of channel assignment and power allocation in each time slot, which leads to the following optimization problem:
%the prioritized SINR, i.e., $\alpha_{k} \gamma_{s, k}^{null}$, is independent of each other in every time slot. As a result, we can transform the complicated overall problem into a sequence of parallel subproblems, and only need to focus on the optimization in each time slot. Hence, a greedy but simple way to solve \eqref{eq:model_nullACI} is proposed as follows
\begin{subequations}\label{eq:model_nullACI1}
	\begin{align}
	\max_{\mathbf{A}_{s}, \mathbf{p}_{s}} \quad &\min_k \alpha_{k}  \gamma_{s, k}^{\mathrm{null}} (\mathbf{A}_{s}, \mathbf{p}_{s}) &\\
	\mbox{s.t.}\quad 
	&\sum\limits_{k=1}^{K} a_{s, k}^{n} \leq 1,\quad \forall n\in \mathcal{N},     \label{subeq:model_nullACI1_b}\\
	&\sum\limits_{n=1}^{N} a_{s, k}^{n} = 1,\quad \forall k\in \mathcal{K},     \label{subeq:model_nullACI1_c} \\
	&a_{s, k}^{n} \in\{0,1\}, \quad \forall k \in \mathcal{K}, \forall n \in \mathcal{N},  \label{subeq:model_nullACI1_d}\\
	&\sum\limits_{k=1}^{K} p_{s, k} \leq P_{\max }, \label{subeq:model_nullACI1_e}\\
	&0 \leq p_{s, k}, \quad \forall k \in \mathcal{K},  \label{subeq:model_nullACI1_f}
	\end{align}
\end{subequations}
where $\mathbf{A}_{s}\triangleq\left[\boldsymbol{a}_{s,1}, \boldsymbol{a}_{s,2}, \ldots, \boldsymbol{a}_{s,K}\right]\in \mathbb{R}^{N \times K}$ denotes the channel assignment matrix with $\boldsymbol{a}_{s,k}\triangleq\big[a_{s, k}^{1}, a_{s, k}^{2}, \ldots, a_{s, k}^{N}\big]^{T}$, and $\mathbf{p}_{s}\triangleq\left[{p}_{s,1}, \ldots, {p}_{s,K}\right]^{T}$ represents the power allocation vector at time slot $s$. 
%Note that the joint channel assignment and power allocation for max-min fairness. 
%Note that applying the fixed point algorithm for max-min fairness problem has been widely studied in recent years \cite{Hong}.
%Note that there is a surge of renewed interests in the max-min fairness problem studied extensively in the literature.
%Note that there is an emerging solution for solving the max-min fairness problem in the literature. such as fixed point algorithm \cite{Hong}, 
Next we develop a semi-closed form solution to this problem. Before presenting the solution, we need the following lemma:
\newtheorem{Lemma1}{Lemma}[section]
\begin{Lemma1} \label{eq:lemma1}
	\emph{In each time slot, the prioritized received SINR of all UAVs are equal at the optimal point of problem \eqref{eq:model_nullACI1}, i.e., } 
	\begin{equation}\label{eq:lemma11}
	\alpha_k \gamma_{s, k}^{\mathrm{null}} = \alpha_m \gamma_{s, m}^{\mathrm{null}}, \quad \forall m, k \in \mathcal{K}.
	\end{equation}	
\end{Lemma1}
%Then we have $\alpha_m \gamma_{s, m}^{null} > \gamma_{s}^{*}$, $\forall m$, hence the constructed solution is better than the optimal solution, which is in contradiction with the initial assumption.
\newtheorem{proof}{Proof}[section]
\begin{proof}
	The key observation is that at the optimal solution, all prioritized SINR are equal. We can prove this by contradiction. Assume that at the optimal solution, there is at least one UAV $k$ that has a higher prioritized SINR than that of others. Let $\gamma_s^{\ast}$ define the minimum prioritized SINR at the optimal point. Since $\alpha_k \gamma_{s, k}^{\mathrm{null}}$ is a continuous increasing function in $p_{s, k}$, we can construct a new solution by reducing $p_{s, k}$ while maintaining that $\alpha_k \gamma_{s, k}^{\mathrm{null}} > \gamma_{s}^{*}$, and increasing the uplink transmission power of the GCU for all the other UAVs to improve the prioritized SINR. Then, we have $\alpha_m
	\gamma_{s,m}^{\mathrm{null}}> \gamma_{s}^{*}$, $\forall m$ and the constructed solution is better than the optimal solution, which contradicts with our previous assumption. As a result, at the optimal solution all prioritized received SINR of different UAVs in the particular time slot are equal, and thus we have	
%	This reveals that a larger optimal value of problem \eqref{eq:model_nullACI1} can be achieved with $\alpha_m \gamma_{s, m}^{null} > \gamma_{s}^{*}$, $\forall m$, which contradicts our
%	presumption. Lemma $3.1$ is thus proved. 
	\begin{equation}\label{eq:lemma12}
	\alpha_k \gamma_{s, k}^{\mathrm{null}} =  \overline {\gamma}_{s}, \quad \forall k \in \mathcal{K},
	\end{equation}	
\end{proof}
where $\overline {\gamma}_{s}$ is the common prioritized received SINR for all UAVs at time slot $s$.

Based on Lemma $3.1$, the uplink transmission power allocated to UAV $k$ at time slot $s$ for any given value of the common prioritized received SINR  can be expressed as
%we can obtain the power allocation to UAV $k$ at time slot $s$ with any given common prioritized SINR $\overline {\gamma}_{s}$ as follow
\begin{equation}\label{eq:common_psk}
p_{s, k} = \overline {\gamma}_{s} \frac{\sum\limits_{n=1}^{N} a_{s, k}^{n} \sigma_{s,k,f_n}^2}{\alpha_k\sum\limits_{n=1}^{N} a_{s, k}^{n} c_{s, k} f_n^{-2}}.
\end{equation}
Note that the objective $\min\limits_{k} \alpha_k \gamma^{\mathrm{null}}_{s,k}(\mathbf{A}_s,\mathbf{p}_s)$ is monotonically increasing with respect to the uplink transmission power $\mathbf{p}_s$ at time slot $s$, it can be concluded that the optimal $(\mathbf{A}^{\ast}_s,\mathbf{p}^{\ast}_s)$ that maximizes $\min\limits_{k} \alpha_k \gamma^{\mathrm{null}}_{s,k}(\mathbf{A}_s,\mathbf{p}_s)$ must be a solution to $\sum_{k=1}^K p_{s,k} = P_{\mathrm{max}}$, i.e. the power budget constraint becomes a strict equality. Consequently, we have
\begin{equation}\label{eq:common_SINR}
\overline{\gamma}_{s} = \Bigg(\sum\limits_{k=1}^{K}\frac{\sum\limits_{n=1}^{N} a_{s, k}^{n} \sigma_{s,k,f_n}^2}{\alpha_k\sum\limits_{n=1}^{N} a_{s, k}^{n} c_{s, k} f_n^{-2}}\Bigg)^{-1} P_{\mathrm{max}}.
\end{equation}
%In this work, our objective is to maximize the minimum weighted SINR among all UAVs. As we discussed above, at the optimal solution all UAVs have a common weighted SINR $\overline\gamma$. According to the expressiong in \eqref{eq:common_SINR},
%we can reformulate problem \eqref{eq:model_nullACI1} as an equivalent minimization problem as follows 
Substituting \eqref{eq:common_psk}-\eqref{eq:common_SINR} into $\alpha_k\gamma_{s,k}^{\mathrm{null}}$, the joint channel assignment and power allocation problem \eqref{eq:model_nullACI1} reduces to a simple channel assignment problem, which is more amenable to optimization and can be represented as
%Based on the above expression, we substitute the result of $\overline {\gamma}_{s}$ into the formulation of \eqref{eq:model_nullACI1}.
%Then the problem \eqref{eq:model_nullACI1} is equivalent to the following minimization problem, in which we only need to optimize the channel assignment $\mathbf{A}_{s}$, i.e.,
\begin{subequations}\label{eq:model_nullACI2}
	\begin{align}
	\min_{\mathbf{A}_{s}} \quad &  \sum\limits_{k=1}^{K}\frac{\sum\limits_{n=1}^{N} a_{s, k}^{n} \sigma_{s,k,f_n}^2}{\alpha_k\sum\limits_{n=1}^{N} a_{s, k}^{n} c_{s, k} f_n^{-2}} \label{eq:model_nullACI2_a}  \\ 
	\mbox{s.t.}\quad 
	& \text{\eqref{subeq:model_nullACI1_b}, \eqref{subeq:model_nullACI1_c}, \eqref{subeq:model_nullACI1_d}.}
	\end{align}
\end{subequations}
Note that in each time slot, each UAV only occupies one channel. In other words, only one element of $\bm{a}_{s,k}\triangleq[a^1_{s,k},\cdots, a^{N}_{s,k}]^T$ is non-zero. As such, problem \eqref{eq:model_nullACI2} can be further simplified into the following equivalent form:
\begin{subequations}\label{eq:model_nullACI3}
	\begin{align}
	\min_{\mathbf{A}_{s}} \quad &  \sum\limits_{k=1}^{K} \sum\limits_{n=1}^{N} a_{s, k}^{n} \frac{ \sigma_{s, k,f_n}^2}{\alpha_{k} c_{s, k} f_n^{-2}} \label{subeq:model_nullACI3_a}  \\ 
	\mbox{s.t.}\quad 
	& \text{\eqref{subeq:model_nullACI1_b}, \eqref{subeq:model_nullACI1_c}, \eqref{subeq:model_nullACI1_d}}. 
	\end{align}
\end{subequations}

Here, we let $\mathbf{\Phi_{s}}\in \mathbb{R}^{N \times K}$ be the prioritized channel quality (PCQ) matrix at time slot $s$ for a given set of UAVs and channels, whose $(n,k)$-th entry $\phi^n_{s,k}$ stands for the PCQ indicator when UAV $k$ is assigned with channel $n$ at time slot $s$ and can be specialized as
%When the channel gain is small but with large channel noise, the channel priority and PCQ indicator is accordingly low.
%We define a $N$$\times$$K$ prioritized channel quality (PCQ) matrix $\mathbf{\Phi}_{s}$ for a given set of UAVs and channels. The $(n, k)^{th}$ entry in this matrix, denated by $\phi_{s, k}^{n}$, represents the PCQ indicator when UAV $k$ accesses the channel $n$, i.e. 
\begin{equation}\label{eq:PCQ_indicator}
[\mathbf{\Phi}_{s}]_{n, k} = \phi_{s, k}^{n} = \frac{ \sigma_{s, k,f_n}^2}{\alpha_{k} c_{s, k} f_n^{-2}}.
\end{equation}
Specifically, $\phi_{s,k}^n$ will be used here as a measure of the preference given by the network to UAV $k$ in assignment of channel $n$. That is, a UAV with a smaller value of parameter $\phi_{s,k}^n$ will have higher priority to be allotted with channel $n$. Considering \eqref{eq:PCQ_indicator}, a UAV with a higher priority $\alpha_k$, larger channel power gain $c_{s,k}$ or smaller EI variance $\sigma^{2}_{s,k,f_n}$ will be given a higher preference to occupy channel $n$. Using the above notations, the objective function in \eqref{eq:model_nullACI3} can be equivalently represented as $\min\limits_{\mathbf{A}_{s}} \sum_{k=1}^{K} \sum_{n=1}^{N} a_{s, k}^{n} \phi_{s, k}^{n}$. It is not difficult to see that the problem \eqref{eq:model_nullACI3} is a linear assignment problem, which can be efficiently solved in polynomial time by using Hungarian algorithm \cite{Hungarian}. The main idea of Hungarian algorithm is to manipulate the objective matrix by adding or subtracting the elements of each row or column until there is at least one zero element in different rows and columns, and then the optimal channel assignment strategy is determined according to the location of zero elements. Specifically, the implementation details of Hungarian algorithm for solving problem \eqref{eq:model_nullACI3} are elaborated below:
%Apparently, when low-prioritized user accesses some channel with small gain but large noise power, the corresponding PCQ indictor will be high. Now the objective function of \eqref{eq:model_nullACI3} can be expressed as $\min\limits_{\mathbf{A}_{s}} \sum_{k=1}^{K} \sum_{n=1}^{N} a_{s, k}^{n} \phi_{s, k}^{n}$. It is easy to see that the problem \eqref{eq:model_nullACI3} is a linear assignment problem, which can be solved in polynomial time by using Hungarian algorithm. The main idea of Hungarian algorithm is to convert the objective matrix by adding or subtracting the elements of each row or column until there is at least one zero element in different rows and columns, and then the optimal assignment method is determined according to the location of zero elements. Specifically, the implementation details of Hungarian algorithm with PCQ matrix $\mathbf{\Phi}_{s}$ are elaborated below
%\noindent Step 1: Row transformation: for each row $n=1,2, \ldots,N$ of $\mathbf{\Phi}_{s}$, $\phi_{s, k}^{n}=\phi_{s, k}^{n}-\min \left\{\phi_{s, k}^{n}| n=1,2, \ldots,N\right\}$.
%\noindent Step 2: Column transformation: for each column $k=1,2, \ldots,K$ of $\mathbf{\Phi}_{s}$, $\phi_{s, k}^{n}=\phi_{s, k}^{n}-\min \left\{\phi_{s, k}^{n}| k=1,2, \ldots,K\right\}$. 	[label = \textbf{\arabic*}]
%\begin{enumerate}\bfseries
%	\item asda
%	\item 12223123
%\end{enumerate}
%\begin{enumerate}[label=\textbf{Step \arabic*:}]
%	\item first thing to do
%	\item Second thing to do
%	\item Third thing to do
%\end{enumerate}

\begin{enumerate}[\textbf{Step}]
	\item \textbf{1:} Row reduction: for each row $n=1,2, \ldots,N$ of PCQ matrix $\mathbf{\Phi}_{s}$, $\phi_{s, k}^{n}=\phi_{s, k}^{n}-\min \left\{\phi_{s, k}^{n}| n=1,2, \ldots,N\right\}$.
	\item \textbf{2:} Column reduction: for each column $k=1,2, \ldots,K$ of $\mathbf{\Phi}_{s}$, $\phi_{s, k}^{n}=\phi_{s, k}^{n}-\min \left\{\phi_{s, k}^{n}| k=1,2, \ldots,K\right\}$. 
	\item \textbf{3:} Check whether the optimal channel assignment strategy can be achieved by covering all zero elements with a minimum number of vertical and horizontal lines. If the number of lines is equal to the order of $\mathbf{\Phi}_{s}$, an optimal set of assignment is obtained, and proceed to step 5. Otherwise, go to next step.
	\item \textbf{4:} If the number of lines needed to cover zero elements is less than the order of $\mathbf{\Phi}_{s}$, transform $\mathbf{\Phi}_{s}$ in the following way:
	\begin{enumerate}[ a.] 		 
		\item Subtracts the minimum element of each row from the uncovered rows.		
		\item Add the minimum element of each column from the covered columns.  
    \end{enumerate}	 	
\end{enumerate}
Repeat Steps 3 and 4 until an optimal set of channel assignment is obtained.

\noindent\textbf{Step 5:} Start with the simple case (a row or column with only one zero element), and cross out both the row and column involved after this channel assignment is finished. As a result, the position of the zero element at the intersection of the crossed out rows and columns is the corresponding optimal channel assignment strategy.
Then continue to perform channel assignment to the remaining rows and columns, with preference to such row or column that has fewer zeros. Repeat the process until all the rows and columns have been crossed out. Finally, the optimal channel assignment matrix $\mathbf{A}_{s}$ is produced.
%\begin{enumerate}[\textbf{Step}]
%	\item \textbf{5:} Start with the simple case (a row or column with only one zero element), and cross out both the row and column involved after this channel assignment is finished. As a result, the position of the zero element at the intersection of the crossed out rows and columns is the corresponding optimal channel assignment strategy.
%	Then continue to perform channel assignment to the remaining rows and columns, with preference to such row or column that has fewer zeros. Repeat the process until all the rows and columns have been crossed out. Finally, the optimal channel assignment matrix $\mathbf{A}_{s}$ is produced.
%\end{enumerate}

After obtaining the optimal channel assignment $a_{s, k}^{n,*}$, it immediately follows from \eqref{eq:common_SINR} that the optimal common prioritized received SINR for all UAVs at time slot $s$ can be specialized as
\begin{equation}\label{eq:optimal_common_SINR}
\overline\gamma_{s} = \bigg[\sum_{k=1}^{K} \frac{ \sigma_{s, k,f_{n}^{*}}^2}{\alpha_k c_{s, k} (f_{n}^{*})^{-2}}\bigg]^{-1} P_{\mathrm{max}}.
\end{equation}
Plugging \eqref{eq:optimal_common_SINR} into \eqref{eq:common_psk}, we can obtain the corresponding uplink transmission power allocated to UAV $k$ at time slot $s$, which is given by
\begin{equation}
p_{s, k}^{*} = \overline\gamma_{s} \frac{ \sigma_{s, k,f_{n}^{*}}^2}{\alpha_k c_{s, k} (f_{n}^{*})^{-2}}.
\end{equation}

%After obtaining the optimal channel assignment $a_{s, k}^{n^{*}}$, it is easy to compute the optimal weighted SINR by $\overline\gamma = P_{max}/\sum_{k=1}^{K} \frac{ \sigma_{s, k,f_{n^{*}}}^2}{\alpha_k c_{s, k} f_{n^{*}}^{-2}}$, as well as the optimal power allocation results $p_{s, k}^{*} = \overline\gamma \frac{ \sigma_{s, k,f_n^{*}}^2}{\alpha_k c_{s, k} f_{n^{*}}^{-2}}$.

The basic idea of the proposed Hungarian-based algorithm is to convert the complicated max-min optimization problem into a series of simple matching subproblems that can be easily solved.
%, which is similar to the fixed point algorithm developed in \cite{Hong} but does not require any additional SINR constraint.
It is worth pointing out that this algorithm leads to the globally optimal solution of problem \eqref{eq:model_nullACI3} since the obtained channel assignment strategy is optimally chosen from all possible spectrum access schemes. For any given PCQ matrix $\mathbf{\Phi_s}$ of dimension $N\times K$, the computational complexity of the proposed Hungarian-based algorithm is $\mathcal{O}\left(K^{2} N\right)$. Note that since the number of UAVs in practice is usually less than that of the available communication channels in each time slot, i.e., $K<N$, the proposed Hungarian-based algorithm can achieve a optimal performance without excessive computational complexity.

%For our input $N$$\times$$K$ prioritized channel quality matrix $\mathbf{\Phi}_{s}$ ($K<N$), the computational complexity of the proposed Hungarian-based algorithm is $\mathcal{O}\left(K^{2} N\right)$.

%which is a weighted bipartite matching problem and can be efﬁciently addressed by the Hungarian method in polynomial time [18].

%In summary, the proposed resource allocation is globally optimal since the reusing pattern is optimally chosen from all possible reusing pairs based on their optimal power allocations. We now discuss the complexity of the proposed method. First, the complexity of the power allocation for all KM possible CUE-DUE pairs is O(KM log(1/ǫ)), where ǫ is the error tolerance of the bisection searching. Second, the Hungarian method solves the spectrum reusing pattern optimization problem in O(M3) time. As a result, the overall complexity of the proposed algorithm is O(KM log(1/ǫ) + M3).

%It is worth pointing out that the proposed method leads to the globally optimal solution since the derived spectrum reusing combination is optimally chosen from all possible reusing patterns based on their optimal power allocations
\newcounter{TempEqCnt2} % 创建临时变量TempEqCnt
\setcounter{TempEqCnt2}{\value{equation}} % 将当前公式序号 赋给TempEqCnt
\setcounter{equation}{20} % 当前公式序号变为x，x等于长公式应有的序号减1.
\begin{figure*}[b]
	\setlength\abovedisplayskip{1pt}
	\setlength\belowdisplayskip{0.5pt}
	\hrulefill % 在公式之前加上一条分割线，长度为页面宽度
	\begin{equation}\label{eq:SINR1_matrix_model}
	\gamma_{s, k}^{\mathrm{A}} (\mathbf{A}, \mathbf{P}) {=} \frac{\mathbf{p}_s^T\bm{e}_k\mathbf{c}_s^T\bm{e}_k\mathbf{f}^T \mathbf{A}_{s} \bm{e}_{k}}{\sum\limits_{ m\neq k}\!\! \bm{e}_k^T\mathbf{A}_s^T\mathbf{W}\mathbf{A}_s\bm{e}_m \mathbf{p}_s^T\bm{e}_m\mathbf{c}_s^T\bm{e}_k\mathbf{f}^T \mathbf{A}_{s} \bm{e}_{m}{+}\bm{e}_k^T\mathbf{\Sigma}_s \mathbf{A}_s\bm{e}_k}.  
	\end{equation}
\end{figure*}
\setcounter{equation}{\value{TempEqCnt1}} % 把TempEqCnt中存的公式序号赋回给当前
\section{EFFICIENT RESOURCE ALLOCATION FOR ACI SYSTEM}
In the previous section, we investigate a special case without consideration of ACI in the uplink multi-UAVs communication system, which greatly simplifies the optimization problem. In this section, we investigate a general scenario in the presence of the considerable ACI due to the limited hardware capability of UAV, where the channel assignment and power allocation are jointly optimized to combat the interference. Specifically, we first transform problem \eqref{eq:SINR2_model} into a compact yet more enunciable matrix form, and subsequently propose a powerful iterative algorithm for its solution based on smooth approximation and alternating optimization methods.

\subsection{Problem Transformation}
For ease of exposition, we introduce some shorthand notations to rewrite problem \eqref{eq:SINR2_model} into a matrix form. The aggregate EI matrix is denoted by $\boldsymbol{\Sigma}_{s}\triangleq\big[\boldsymbol{\sigma}_{s,1}, \boldsymbol{\sigma}_{s,2}, \ldots, \boldsymbol{\sigma}_{s,K}\big]^{T}\in \mathbb{R}^{K \times N}$, where $\boldsymbol{\sigma}_{s,k}\triangleq\big[\sigma_{s,k, f_1}^{2}, \sigma_{s,k,f_2}^{2},\ldots, \sigma_{s, k,f_N}^{2}\big]^{T}$ represents the power of EI on each channel of UAV $k$ at time slot $s$. Furthermore, we collect the ACI coefficients in a symmetric matrix $\mathbf{W}\in \mathbb{R}^{N \times N}$, where the $(i,j)$-th entry of $\mathbf{W}$ is given by ${W_{i, j} \triangleq \mu_{{f_{i}}{f_{j}}}}$. In addition, let $\bm{c}_s\triangleq [c_{s,1},\cdots,c_{s,K}]^T$ denote the composite channel gain vector between the GCU and UAV swarm, and $\bm{f}\triangleq [f_1^{-2},\cdots,f_N^{-2}]^T$.

%\newtheorem{Proposition1}{Proposition}[section]
%\begin{Proposition1} \label{eq:Proposition1}
%	\emph{Problem \eqref{eq:SINR2_model} can be equivalently transformed as  \eqref{eq:MMF_ACI}}, as shown at the bottom of the page, where $\bm{e}_k$ is an $M$-dimensional unit column vector with the $k$-th element being $1$.
%\end{Proposition1}
%
%\newtheorem{proof1}{Proof}[section]
%\begin{proof}
%	
%\end{proof}
%Based on the above definition, we have the following proposition.

Now we are ready to rewrite \eqref{eq:SINR1_model}. First, denote by $\bm{e}_k$  an $K$-dimensional unit column vector with the $k$-th element being $1$. Then by the definitions of $\mathbf{\Sigma}_s$ and $\mathbf{A}_s$, we clearly have  $\bm{\sigma}_{s,k}=\mathbf{\Sigma}_s^T \bm{e}_k$ and $\bm{a}_{s,k}=\mathbf{A}_s\bm{e}_k$. Recall from \eqref{eq:constrain1}-\eqref{eq:constrain3} that each UAV is assigned with only one channel at time slot $s$ (i.e., $\bm{a}_{s,k} \in \left\{\bm{e}_1,\cdots,\bm{e}_K\right\}$), we can further obtain 
%\begin{center}
%\begin{subequations}
%	\begin{align*}
%	 & \sum\limits_{n=1}^{N} a_{s, k}^{n} f_{n}^{-2} = \mathbf{f}^T\bm{a}_{s,k} = \mathbf{f}^T \mathbf{A}_{s} \boldsymbol{e}_{k}, \\
%	& \sum\limits_{n=1}^{N} \sum\limits_{j=1}^{N} a_{s, k}^{n} a_{s, m}^{j} \mu_{f_{n},f_{j}}=\bm{a}_{s,k}^T\mathbf{W}\bm{a}_{s,m}=\boldsymbol{e}_{k}^{T} \mathbf{A}_{s}^{T} \mathbf{W} \mathbf{A}_{s} \boldsymbol{e}_{m}, \\
%	& \sum\limits_{n=1}^{N} a_{s, k}^{n} \sigma_{s,k,f_{n}}^2=\bm{\sigma}_{s,k}^T\bm{a}_{s,k}= \bm{e}_k^T\mathbf{\Sigma}_s \mathbf{A}_s\bm{e}_k.
%	\end{align*}
%\end{subequations}
%\end{center}
\begin{subequations}
	\setlength\abovedisplayskip{3pt}
	\setlength\belowdisplayskip{1pt}
	\begin{align*}
	\sum\limits_{n=1}^{N} a_{s, k}^{n} f_{n}^{-2} = \mathbf{f}^T\bm{a}_{s,k} = \mathbf{f}^T \mathbf{A}_{s} \boldsymbol{e}_{k},
	\end{align*}
\end{subequations}
\begin{subequations}
	\setlength\abovedisplayskip{3pt}
	\setlength\belowdisplayskip{1pt}
	\begin{align*}
	\sum\limits_{n=1}^{N} \sum\limits_{j=1}^{N} a_{s, k}^{n} a_{s, m}^{j} \mu_{f_{n},f_{j}}=\bm{a}_{s,k}^T\mathbf{W}\bm{a}_{s,m}=\boldsymbol{e}_{k}^{T} \mathbf{A}_{s}^{T} \mathbf{W} \mathbf{A}_{s} \boldsymbol{e}_{m},
	\end{align*}
\end{subequations}
\begin{subequations}
	\setlength\abovedisplayskip{3pt}
	\setlength\belowdisplayskip{1pt}
	\begin{align*}
	\sum\limits_{n=1}^{N} a_{s, k}^{n} \sigma_{s,k,f_{n}}^2=\bm{\sigma}_{s,k}^T\bm{a}_{s,k}= \bm{e}_k^T\mathbf{\Sigma}_s \mathbf{A}_s\bm{e}_k.
	\end{align*}
\end{subequations}
		
By using the above three identities and further noting ${p}_{s,k}={\mathbf{p}_{s}^{T}} \boldsymbol{e}_{k}$ and $c_{s,k} = {\mathbf{c}_{s}^{T}} \boldsymbol{e}_{k}$, $\forall k$, \eqref{eq:SINR1_model} can be rewritten as $\gamma^{\mathrm{A}}_{s,k}$ at the bottom of this page. Consequently, problem \eqref{eq:SINR2_model} can be equivalently rewritten as follows  
\setcounter{equation}{21}  
\begin{subequations}\label{eq:MMF_ACI}
	\begin{align}
	\max_{\mathbf{A}, \mathbf{P}} \quad & \min _{k} \frac{1}{S}\sum_{s=1}^S
	\alpha_k\gamma_{s, k}^{\mathrm{A}}(\mathbf{A}, \mathbf{P})
	& \label{subeq:MMF_ACI_a} \\
	\mbox{s.t.}\quad 
	&\mathbf{A}_{s} \bm{e}_{k} \neq \mathbf{A}_{s} \bm{e}_{m}, \quad \forall k,m\in \mathcal{K}, \quad \forall s \in \mathcal{S}, \label{subeq:MMF_ACI_b} \\
	&\mathbf{1}^{T} \mathbf{A}_{s} \boldsymbol{e}_{k}=1, \quad \forall k \in \mathcal{K}, \quad \forall s \in \mathcal{S}, \label{subeq:MMF_ACI_c} \\
	&a_{s, k}^{n} \in\{0,1\}, \quad \forall k \in \mathcal{K},  \forall n \in \mathcal{N},  \forall s \in \mathcal{S},  \label{subeq:MMF_ACI_d} \\
	&\mathbf{1}^{T} \mathbf{p}_{s} \leq P_{\max }, \quad \forall s \in \mathcal{S}, \label{subeq:MMF_ACI_e} \\
	&0 \leq p_{s, k}, \quad \forall k \in \mathcal{K}, \quad \forall s \in \mathcal{S}. \label{subeq:MMF_ACI_f}
	\end{align}
\end{subequations}

As discussed in Section III, we can decompose problem \eqref{eq:MMF_ACI} into $S$ independent subproblem across different time slot. Hence, at each time slot $s$, we only need to solve the following max-min fairness problem
\setcounter{equation}{22}  
\begin{subequations}\label{eq:MMF_ACI_problem}
	\begin{align}
	\max_{\mathbf{A}_s, \mathbf{p}_s} \quad & \min_k \alpha_k \gamma_{s,k}^{\mathrm{A}}(\mathbf{A}_s, \mathbf{p}_s)  \\ 
	\mbox{s.t.}\quad 
	&\mathbf{A}_{s} \bm{e}_{k} \neq \mathbf{A}_{s} \bm{e}_{m},\quad \forall k,m\in \mathcal{K},   \label{subeq:23_b}\\
	&\mathbf{1}^{T} \mathbf{A}_{s} \boldsymbol{e}_{k}=1, \quad \forall k \in \mathcal{K},    \label{subeq:23_c}\\
	&a_{s, k}^{n} \in\{0,1\}, \quad \forall k \in \mathcal{K}, \quad \forall n \in \mathcal{N},  \label{subeq:23_d} \\
	&\mathbf{1}^{T} \mathbf{p}_{s} \leq P_{\max },\\
	&0 \leq p_{s, k}, \quad \forall k \in \mathcal{K}.
	\end{align}
\end{subequations}
Note that the problem \eqref{eq:MMF_ACI_problem} is still a MINP problem. However, it can be observed that the constraints are separable with respect to the optimization variables $\mathbf{A}_s$ and $\mathbf{p}_s$. In what follows, we propose an iterative algorithm to solve problem \eqref{eq:MMF_ACI_problem} based on alternating optimization (AO). Specifically, the proposed algorithm alternates between two steps until convergence. In step 1, fixing $\mathbf{p}_s$ we update $\mathbf{A}_s$ by using gradient projection (GP) method. And in step 2, fixing  $\mathbf{A}_s$ we update  $\mathbf{p}_s$ by eigenvalue decomposition method. The details of the proposed algorithm are elaborated as follow.

\subsection{Channel Assignment Optimization}
This subsection is devoted to describing how we perform step 1 in our alternating algorithm, i.e., how to solve the following problem with fixed $\mathbf{p}_s$
\begin{subequations}\label{eq:MMF-As}	
		\setlength\abovedisplayskip{3pt}
		\setlength\belowdisplayskip{1pt}
	\begin{align}
	\max_{\mathbf{A}_s} \quad & \min_k \alpha_k \gamma_{s,k}^{\mathrm{A}}(\mathbf{A}_s, \mathbf{p}_s) \\
	\mbox{s.t.}\quad
	& \text{\eqref{subeq:23_b}, \eqref{subeq:23_c}, \eqref{subeq:23_d}.}
	\end{align}
\end{subequations}
Note that, besides the binary variable constraint \eqref{subeq:23_d}, constraint \eqref{subeq:23_b} further makes the above problem intractable. To address the challenge arising from \eqref{subeq:23_b}, we deliberately set the diagonal elements of the matrix $\mathbf{W}$ to be sufficiently large (denoted by $\hat{\mathbf{W}}$) so as to increase the penalty when the same channel is simultaneously assigned to two different UAVs. It is seen that as the diagonal elements of $\mathbf{W}$ are getting large, the term $\bm{e}_k^T\mathbf{A}_s^T\mathbf{W}\mathbf{A}_s\bm{e}_m$ will drastically increase when $\bm{e}_k = \bm{e}_m$, which forces UAVs to access different channels for achieving higher SINR. In this way, we can relax \eqref{subeq:23_b} without performance loss. 

To make problem \eqref{eq:MMF-As} more tractable, we relax the binary variables in \eqref{subeq:23_d} to the box constraint $a_{s, k}^{n} \in[0, 1]$. Ultimately, it can be obtained by rounding the optimization results into 0-1 integers. 
However, we find from simulations that this relaxation often leads to an undesirable numerical solution (i.e., roughly equals 0.5). Under this setup, naive rounding scheme would inevitably incur huge performance loss. 
In addition, we also observe some counter-intuitive outcomes that a smaller value of $a^n_{s,k}$ unexpectedly yields a larger SINR $\gamma^{\mathrm{A}}_{s,k}$, which implies that UAV $k$ without occupancy of  channel $n$ at time slot $s$ achieve better SINR performance instead. 
The reason for this weird phenomenon is that the numerator of SINR $\gamma^{\mathrm{A}}_{s,k}$  has smaller order in $\mathbf{A}_s$ than that of the ACI term in its denominator, and thus the denominator of SINR $\gamma^{\mathrm{A}}_{s,k}$ will suffer from faster attenuation as compared to its numerator when  the value of $a^n_{s,k}$ decreases.
To address these above challenges, we  introduce an additional exponent to the term $\bm{e}^T_k\mathbf{A}^T_s
\mathbf{A}_s\bm{e}_k$ in the numerator of SINR $\gamma^{\mathrm{A}}_{s,k}$ while not changing the physical meaning of \eqref{eq:MMF-As}. Recall from \eqref{subeq:23_d} that $a^n_{s,k}$ is a binary variable before relaxation, we have  
\begin{equation}
(\bm{e}_{k}^{T} \mathbf{A}_{s}^{T} \mathbf{A}_{s} \bm{e}_{k})^{\tau} = \bm{e}_{k}^{T} \mathbf{A}_{s}^{T} \mathbf{A}_{s} \bm{e}_{k},\quad  \tau\geq 1.
\end{equation}
When $\tau$ is sufficiently large, the numerator of $\gamma_{s,k}^{\mathrm{A}}$ has larger order of $\mathbf{A}_s$ than that of its denominator, consequently the numerator of $\gamma_{s,k}^{\mathrm{A}}$ will diminish at a faster speed as compared to its denominator when the value of $a^n_{s,k}$ decreases.
Accordingly, problem \eqref{eq:MMF-As} is finally relaxed to the following
\begin{subequations}\label{eq:MMF-As2}
	\setlength\abovedisplayskip{3pt}
	\setlength\belowdisplayskip{1pt}
	\begin{align}
	\max_{\mathbf{A}_s} \quad & \min_k \hat{f}_{s,k}(\mathbf{A}_s) \label{subeq: MMF-As2-A} \\
	\mbox{s.t.}\quad 
	&\mathbf{1}^{T} \mathbf{A}_{s} \boldsymbol{e}_{k}=1, \quad \forall k \in \mathcal{K},  \\
	&0\leq a_{s, k}^{n}\leq 1, \quad \forall k \in \mathcal{K}, \quad \forall n \in \mathcal{N},%\label{subeq: 4d}
	\end{align}
\end{subequations}
where 
\begin{small}	
\begin{align*}
\hat{f}_{s,k}(\mathbf{A}_{s}) {\triangleq} \frac{\alpha_{k} {p}_{s,k}\mathbf{c}_s^T\bm{e}_k\mathbf{f}^T \mathbf{A}_{s} \bm{e}_{k} \left(\bm{e}_{k}^{T} \mathbf{A}_{s}^{T} \mathbf{A}_{s} \bm{e}_{k} \right)^{\mathrm{\tau}}}{\sum\limits_{m\neq k}\!\! \bm{e}_k^T\mathbf{A}_s^T\hat{\mathbf{W}}\mathbf{A}_s\bm{e}_m {p}_{s,m}\mathbf{c}_s^T\bm{e}_k\mathbf{f}^T \mathbf{A}_{s} \bm{e}_{m}{\!+}\bm{e}_k^T\!\mathbf{\Sigma}_s \mathbf{A}_s\bm{e}_k}
\end{align*}
\end{small}

\noindent with $\tau$ being some integer. The earlier work in \cite{relate_xue} proves that the algorithm can achieve better numerical results with properly selected $\tau$, which is also verified by the simulation in Section V.

Now the difficulty remains to be in the non-smooth part of the objective function in \eqref{subeq: MMF-As2-A}. To address this issue, we hereby apply the smooth approximation to the objective of problem \eqref{eq:MMF-As2}, using the following lemma. 

\newtheorem{Lemma2}{Lemma}[section]
\begin{Lemma2}  \label{eq:lemma2}
	\emph{According to the smooth approximation method in \cite{smooth}, the following inequality holds true}
	\begin{equation}
	\max \left\{x_{1}, \ldots, x_{K}\right\}\leq {f}(x)\leq\max \left\{x_{1}, \ldots, x_{K}\right\} + \mu\log K,
	\end{equation}
	\emph{where}
	\begin{equation}
	{f}(x) {=}  \mu \log \left(\sum_{k=1}^K\exp \left(\frac{x_{k}}{\mu}\right)\right),
	\end{equation}
	\emph{along with a sufficiently small positive constant $\mu>0$. Then we have}
	\begin{equation}
	{f}(x) \approx  \max \left\{x_{1}, \ldots, x_{K}\right\}.
	\end{equation}
\end{Lemma2}

According to Lemma $4.1$, we have the following approximation 
%Now the difficulty remains to be in the nonsmooth part of the objective function. To address this issue, we develop an iterative algorithm to address problem \eqref{eq:MMF-As2} by using gradient projection (GP) method combined with a smoothing technique. According to the smooth approximation method in \cite{b10}, we have the following approximation 
$$-\min \left\{x_{1}, \ldots, x_{K}\right\}{=}\mu \log \left(\sum_{k=1}^K\exp \left(\frac{-x_{k}}{\mu}\right)\right),$$
Therefore, we have
%\begin{align}\label{eq:smooth}
% f_{s, \mu}(\mathbf{A}_{s}) &\triangleq\mu \log \sum_{k=1}^{K} \exp \left(\frac{-\hat{f}_{s, k}(\mathbf{A}_{s})}{\mu}\right)\\
%& \approx -\min \left\{\hat{f}_{s, 1}(\mathbf{A}_{s}), \ldots, \hat{f}_{s, K}(\mathbf{A}_{s})\right\}. 
%\end{align}
\begin{equation}
\begin{aligned}\label{eq:smooth}
f_{s, \mu}\left(\mathbf{A}_{s}\right) & \triangleq \mu \log \sum_{k=1}^{K} \exp \left(\frac{-\hat{f}_{s, k}\left(\mathbf{A}_{s}\right)}{\mu}\right) \\
& \approx-\min \left\{\hat{f}_{s, 1}\left(\mathbf{A}_{s}\right), \ldots, \hat{f}_{s, K}\left(\mathbf{A}_{s}\right)\right\}.
\end{aligned}
\end{equation}

Using \eqref{eq:smooth}, we can approximate problem \eqref{eq:MMF-As2} as follows
\begin{subequations}\label{eq:MMF-As3}
	\begin{align}
	\min_{\mathbf{A}_s} \quad & f_{s, \mu}(\mathbf{A}_{s}) \\
	\mbox{s.t.}\quad
	&\mathbf{1}^{T} \mathbf{A}_{s} \boldsymbol{e}_{k}=1, \quad \forall k \in \mathcal{K},  \\
	&0\leq a_{s, k}^{n}\leq 1, \quad \forall k \in \mathcal{K}, \quad \forall n \in \mathcal{N}.
	\end{align}
\end{subequations}

\begin{algorithm}[!t]
	\caption{Proposed Gradient Projection Algorithm for problem \eqref{eq:MMF-As3}}
	\label{algGP}
	\begin{algorithmic}[1]
		\STATE \textbf{Initialize} $\mathbf{A}_{s}$. Define the tolerance of accuracy $\epsilon_3$ and the maximum iteration number $T_{1,\max}$.
		\REPEAT
		\STATE Calculate the gradient to obtain the temporary solution:
		\begin{center}
			 $\bar{\mathbf{A}}_s = \mathbf{A}_{s} - \nabla_{\mathbf{A}_{s}} f_{s, \mu}(\mathbf{A}_{s})$
			\end{center}
		\STATE Calculate the projections:
		\begin{center}
			${\mathbf{A}_{s}^\mathrm{proj}}=P_{\Omega_{\mathbf{A}_{s}}}\left(\bar{\mathbf{A}}_s\right)$
		\end{center}	
		\STATE Update $\mathbf{A}_{s}$ according to the following
		\begin{center}
			$\mathbf{A}_{s} \longleftarrow \mathbf{A}_{s}+\alpha\left({\mathbf{A}_{s}^\mathrm{proj}}-\mathbf{A}_{s}\right)$
		\end{center}	
		\UNTIL{the objective value of $f_{s, \mu}(\mathbf{A}_{s})$ converges, or the maximum iteration number is reached.}	
		\STATE Round the results of $\mathbf{A}_{s}$  into 0-1 integers.	
	\end{algorithmic}
\end{algorithm}

%To address problem \eqref{eq:MMF-As3}, we develop an iterative algorithm by using gradient projection (GP) method combined with the smooth approximation.
Note that since the projection onto the simple constraint space can be easily calculated \cite{GP,xihan1,xihan2}, we hereby apply the GP method to solve problem, mainly consisting of three computational steps. Specifically, the first step is to obtain a temporary solution $\bar{\mathbf{A}}_s$ by calculating the gradient of function $f_{s,\mu}(\mathbf{A}_s)$ with respect to $\mathbf{A}_s$, followed by projecting this solution $\bar{\mathbf{A}}_s$ onto the constraint space $\Omega_{\mathbf{A}_s}$ of problem \eqref{eq:MMF-As3} to derive the next solution in the second step. In the third step, we obtain the optimal solution $\mathbf{A}^{\ast}_s$ by search in the opposite direction $\mathbf{A}^{\mathrm{proj}}_s-\mathbf{A}_s$ with a properly chosen step size $\alpha$. 
The detailed update procedure are summarized in Algorithm \ref{algGP}, where  $P_{\Omega_{\mathbf{A}_{s}}}(\mathbf{X})$ refers to the projection of $\mathbf{X}$ onto $\Omega_{\mathbf{A}_{s}}$. Moreover, the convergence characteristics of the proposed GP method crucially depend on the choice of step size $\alpha$. In this paper, we adopt the modified Armijo step size rule as suggested in \cite{Arimijo,xihan3,xihan4}. 
With properly-chosen step size $\alpha$, the objective function value of $f_{s,\mu}(\mathbf{A}_s)$ is guaranteed to monotionically decrease through the iterations from Line 3 to Line 5 in Algorithm 1. 
Once the iterations are finished, the optimal channel assignment $\mathbf{A}_{s}^{*}$ can be easily obtained by respectively rounding their elements into $0$-$1$ integers.

\newtheorem{remark}{Remark}
\begin{remark}  
%	To gain more insights, it is worth further considering a special problem that UAV should not switch channel frequently in order to save energy. As the optimal channel assignment $\mathbf{A}_{s}^{*}$ at time slot $s$ obtained, we can account for this problem just simply by observing whether or not a significant min-fairness improvement can be achieved by changing the channel assignment scheme from $\mathbf{A}_{s-1}^{*}$ to $\mathbf{A}_{s}^{*}$ ($s > 1$). For example, we set a specific handover threshold $\theta$. If a $\theta$ (e.g., 20\%) min-fairness improvement is made, then perform channel handover; otherwise, just keep the channel assignment scheme unchanged. In this way, the frequency of channel handovers can be greatly reduced.
	In practice, UAV usually does not switch channel frequently in consideration of energy saving and signaling overhead, thereby prolonging the endurance of UAV systems. Toward this end, we can change the channel assignment scheme from $\mathbf{A}^{\ast}_{s-1}$ to $\mathbf{A}^{\ast}_s$ by simply observing whether the corresponding minimum SINR improvement is larger than a predefined handover threshold $\theta$. For example, we set $\theta=20\%$ and perform channel handover if a $30\%$ minimum SINR improvement is made. Otherwise, the channel assignment scheme remains invariant within time slot $s$. In practice, the  predefined handover threshold $\theta$ should be appropriately tuned to achieve a favorable trade-off between performance improvement and signaling overhead.
	
	%To gain more insights, it is worth further considering a special problem that UAV should not switch channel frequently in order to save energy (thus named non-frequent channel-handover objective). As obtained optimal channel assignment $\mathbf{A}_{s}^{*}$ in \eqref{eq:MMF-As3}, we can account for the non-frequent channel-handover objective just simply by observing whether or not a significant min-fairness improvement can be achieved by changing the channel assignment scheme from $\mathbf{A}_{s-1}^{*}$ to $\mathbf{A}_{s}^{*}$.
\end{remark}

\subsection{Power Allocation Optimization}
In this subsection, we show how to solve problem \eqref{eq:MMF_ACI_problem} for $\mathbf{p}_s$ while fixing $\mathbf{A}_{s}$. For given $\mathbf{A}_{s}$, problem \eqref{eq:MMF_ACI_problem} is reduced to
\begin{subequations}\label{eq:MMF-Ps}
		\setlength\abovedisplayskip{3pt}
	\setlength\belowdisplayskip{1pt}
\begin{alignat}{2}
\max\limits_{ \mathbf{p}_s } ~
& \min\limits_k \left\{ \frac{p_{s,k} b_{s,k,k}} {\sum\limits_{m \neq k}^K p_{s,m} b_{s,k,m} + \sigma_{s,k}^2  }\right\}   \label{subeq: MMF-Psa} \\ 
\text{s.t.}  \quad
& \text{\eqref{subeq:model_nullACI1_e}, \eqref{subeq:model_nullACI1_f}.}
\end{alignat}
\end{subequations}
where $b_{s,k,m}$'s are all constants, which can be easily calculated based on \eqref{eq:SINR1_matrix_model} with the fixed $\mathbf{A}_{s}$. 
Following the similar approach in \cite{Optimal_downlink_eig}, we can globally solve the above problem by resorting to eigenvalue decomposition. Specifically, let $\delta_s\triangleq \frac{p_{s,k} b_{s,k,k}} {\sum\limits_{m \neq k}^K p_{s,m} b_{s,k,m} + \sigma_{s,k}^2  }$ be the objective value in \eqref{subeq: MMF-Psa} and define $\bm{z}_{s} \triangleq  [p_{s,1},\ldots,p_{s,K}, 1]^T$,
%$\eta_{s}$ denote the objective value  $\frac{p_{s,k} b_{s,k,k}} {\sum\limits_{m \neq k}^K p_{s,m} b_{s,k,m} + \sigma_{s,k}^2  }$, define $\bm{z}_{s} \triangleq  [p_{s,1},\ldots,p_{s,K}, 1]^T$, and 
\begin{equation*}
\mathbf{C}_{s} \triangleq
\begin{aligned}
\begin{bmatrix}
\mathbf{I}_{K \times K} & \mathbf{0}_{K \times 1} \\
\mathbf{1}_{1 \times K} & - P_{\max}
\end{bmatrix}
\end{aligned} ,\quad
\mathbf{B}_{s} \triangleq
\begin{aligned}
\begin{bmatrix}
\mathbf{{R}}_{s, K \times K} & {\bm{h}_{s}} \\
\mathbf{0}_{1 \times K} & 0
\end{bmatrix}
\end{aligned} ,
\end{equation*}
with
\begin{equation*}
\mathbf{R}_{s} = \left\{
\begin{aligned}
\frac{b_{s,k,m}}{b_{s,k,k}} , \quad & k \neq m \\
0~~,\quad & k = m
\end{aligned},
\right.
\quad\bm{h}_{s} = \left[
\frac{\sigma_{s,1}^2}{b_{s,1,1}}, \ldots,
\frac{\sigma_{s,K}^2}{b_{s,K,K}}
\right]^T.
\end{equation*}
%Note that the objective $\min\limits_{k} \alpha_k \gamma^{\mathrm{null}}_{s,k}(\mathbf{A}_s,\mathbf{p}_s)$ is monotonically increasing with respect to the uplink transmission power $\mathbf{p}_s$ at time slot $s$, we can conclude that maximizes $\min\limits_{k} \alpha_k \gamma^{\mathrm{null}}_{s,k}(\mathbf{A}_s,\mathbf{p}_s)$ must be a solution to $\sum_{k=1}^K p_{s,k} = P_{\mathrm{max}}$, i.e. the power budget constraint becomes a strict equality. Consequently, we have
Note that although \eqref{subeq:model_nullACI1_e} is an inequality constraint, the objective value of \eqref{subeq: MMF-Psa} is monotonically increasing with respect to the uplink transmission power $\mathbf{p}_s$ at time slot $s$, consequently we can conclude that the \eqref{eq:MMF-Ps} must satisfy the solution of $\sum_{k=1}^K p_{s,k} = P_{\mathrm{max}}$, i.e., all the available transmit power of the GCU is applied to deliver useful information at each time slot.
Based on the above discussions, the constraint of problem \eqref{eq:MMF-Ps} can be rewritten as
\begin{equation}
\mathbf{C}_{s} \bm{z}_{s} = \delta_s \mathbf{B}_{s} \bm{z}_{s}.
\end{equation}
Considering that $\mathbf{C}_{s}$ is nonsingular, we further have
\begin{equation}\label{eq:ei}
\frac{1}{\delta_s} \bm{z}_{s} = \mathbf{C}_{s}^{-1} \mathbf{B}_{s} \bm{z}_{s},
\end{equation}
%Here, we have $\frac{1}{{\lambda}_{s}} \bm{z}_{s} = \mathbf{C}_{s}^{-1} \mathbf{B}_{s} \bm{z}_{s}$,
where $\frac{1}{\delta_s}$ is the eigenvalue of the non-negative matrix $\mathbf{C}_{s}^{-1} \mathbf{B}_{s}$, and $\bm{z}_{s}$ is the corresponding eigenvector. 
According to the property of non-negative matrix \cite{Optimal_downlink_eig}, both the largest eigenvalue and its corresponding eigenvector are positive, which indicates that constraint \eqref{subeq:model_nullACI1_e} is automatically satisfied. Based on Theorem 2 in \cite{Optimal_downlink_eig}, it follows that all prioritized SINR of different UAVs are equal at the optimal power allocation scheme, and the value of optimal prioritized SINR is the reciprocal of the largest eigenvalue of $\mathbf{C}^{-1}_s\mathbf{B}_s$. Consequently, we have 
\begin{equation}\label{eq:mu}
\delta_s = \frac{1}{\lambda_{\max}(\mathbf{C}_{s}^{-1} \mathbf{B}_{s})},  
\end{equation}
where $\lambda_{\max}(\cdot)$ denotes the maximum eigenvalue of a matrix. Then we can scale the corresponding eigenvector $\bm{z}_{s}$ so that the last element is normalized to one. Accordingly, the first $K$ elements of $\bm{z}_{s}$ constitute the optimal solution of problem \eqref{eq:MMF-Ps}, denoted by $\mathbf{p}^{\ast}_s$.

%Due to the fact that the goal of us is to maximize the minimum SINR of UAV swarm, so we could find the largest eigenvalue of the matrix $\mathbf{C}_{s}^{-1} \mathbf{B}_{s}$. Obviously, the corresponding eigenvector $\bm{z}_{s}$ can represent the optimal power allocation strategy $\mathbf{p}_{s}$ for UAVs at time slot $s$.

\begin{algorithm}[!t]
	\caption{Proposed Alternating Optimization Algorithm for problem \eqref{eq:MMF_ACI_problem}} \label{algAO}
	\begin{algorithmic}[1]
		\STATE \textbf{Initialize} $\mathbf{p}_{s}^{0}$, $\mathbf{A}_{s}^{0}$, and  define the maximum iteration number $T_{2,\max}$. Set $t_{1} = 0$.		
		\REPEAT
		\STATE Apply Algorithm 1 with input $\mathbf{p}_s^{t_{1}}$ to  $\mathbf{A}^{t_{1}+1}_s$ as elaborated in Section IV-B.
%		\STATE Solve problem \eqref{eq:MMF-As3} using Algorithm 1 for given $\mathbf{p}_{s}^{t}$, and obtain $\mathbf{A}_{s}^{t+1}$.
		\STATE Solve problem \eqref{eq:MMF-Ps} via  eigenvalue decomposition for given $\mathbf{A}_{s}^{t_{1}+1}$, and let $\mathbf{p}^{t_{1}+1}_s=\mathbf{p}^{\ast}_s$.
		\STATE Update $t_{1} = t_{1} + 1.$
		\UNTIL{the objective value converges, or the maximum iteration number is reached.} 
	\end{algorithmic}
\end{algorithm}

\subsection{Overall Description and Computation Complexity}
Based on the above steps, we summarize the proposed AO algorithm in Algorithm \ref{algAO}.
Specifically, in each iteration, the channel assignment $\mathbf{A}_{s}$ and power allocation $\mathbf{p}_{s}$ are alternatingly optimized with the other fixed. As a result, problem \eqref{eq:MMF-As3} or \eqref{eq:MMF-Ps} can be solved correspondingly. Moreover, the initial value of the next iteration is the result of this step, until the objective value converges.

Next, we are devoted to analyzing the computation complexity of the proposed AO algorithm in terms of the number of floating point operations (FPOs). In each iteration of this algorithm, we solve the subproblems for the two blocks of variables sequentially.

\begin{enumerate}[\hspace{1em}1)] 		 
	\item Let us focus on the subproblem with respect to $\mathbf{A}_s$. Notwithstanding the computation of the invariant term, the complexity of updating $\mathbf{A}_s$ is dominated by the gradient calculation of objective function $f_{s,\mu}$ and is given by $\mathcal{O}(T_1N^2K^2)$, where $T_1$ is the number of iterations required by Algorithm 1.	
	\item Next, we turn attention to the subproblem with respect to $\mathbf{p}_s$, which is dominated by two parts. The first part calculates the matrix inversion of $\mathbf{C}_s$ based on the Gaussian Jordan elimination with complexity $\mathcal{O}((K+1)^{3})$. The second part performs the eigenvalue decomposition of  $\mathbf{C}^{-1}_s \mathbf{B}_s$ with complexity of $\mathcal{O}((K+1)^{3})$. Thus the overall computational complexity for updating $\mathbf{p}_s$ is given by $\mathcal{O}((K+1)^{3})$.	
%	calculate the the largest eigenvalue and corresponding eigenvector of the non-negative matrix $\mathbf{C}^{-1}_s\mathbf{B}_s$.	
%	the complexity of obtaining $\mathbf{p}^{\ast}_s$ is mainly dominated by the computation of the largest eigenvalue and corresponding eigenvector of the non-negative matrix $\mathbf{C}^{-1}_s\mathbf{B}_s$, which is given by $ \mathcal{O}\left((K+1)^{3}\right)$.
\end{enumerate}	

Based on the above analysis, the overall computational complexity of the proposed AO algorithm can be expressed as $\mathcal{O}((N^{2}K^{2}T_{1}+(K+1)^{3})T_{2})$, where $T_{2}$ is the number of iterations required by the proposed AO algorithm.

%Note that in each interation, the complexity of the proposed GP algorithm is mainly dominated by the computation of gradient, which has a complexity of $\mathcal{O}\left(N^{2}K^{2}\right)$. Consequently, the computational complexity of the overall algorithm is $\mathcal{O}\left(N^{2}K^{2}T\right)$, where $T$ is the number of iterations required to reach a termination condition. 

%we propose an iterative method to solve problem \eqref{eq:MMF_ACI_problem} by adopting the AO algorithm. Specifically, in each iteration, the channel assignment $\mathbf{A}_{s}$ and power allocation $\mathbf{P_{s}}$ are alternately optimized with the other fixed. Therefore, the problem \eqref{eq:MMF-As3} or \eqref{eq:MMF-Ps} can be solved correspondingly. Moreover, the initial value of the next iteration is the result of this step, until the objective value converges. The detailed process is in Algorithm \ref{algAO}.

\section{NUMERICAL RESULTS}
This section presents numerical simulations to validate the effectiveness of the proposed algorithm and draw some essential insights. 
In the simulation, we consider a multi-UAV enabled mission execution scenario where the UAV swarm flies at the altitude of $H=500$ m with a predefined trajectory and maintains communication with the GCU. 
Furthermore, the location of the GCU and destination are set to $\boldsymbol{x}_{0} = (0, 0, 0)$ and $\boldsymbol{x}_{D} = (1000, 0, 0)$, respectively. 
For simplicity, we assume that the UAV formation keeps the maximum flight speed $V_{\mathrm{max}}=50$ m/s \cite{UAV_speed}, and each time slot is $1$ s. Hence, the whole flight duration for completing the mission is $20$ s.   
The additional pass loss for LoS link and NLoS link are respectively set as $\eta_{L} = \text{3dB}$ and $\eta_{N} = \text{23dB}$.
There are $L$ radiation sources randomly distributed across the region of $2\times2$ k$\mathrm{m}^2$, and their interference power is set to about $\sigma_{s,k,f_n}^2=-10\text{ dBm}$. Moreover, the maximum transmission power of the GCU is set to $P_{\mathrm{max}}=\text{30 dBm}$ \cite{xihan5}, and the available channels are modeled as $f_n=F+n\times \triangle f$, where $F=500$ MHz refers to the baseline carrier frequency and $\triangle f= 5$ MHz represents the channel interval. 
For the ACI coefficients $\mu_{f_{1}, f_{2}}$, we use the existing practical experimental results of \cite{ACI_date} to set ACI coefficients matrix $\mathbf{W}$. For simplicity, the weighted factor of each UAV $\alpha_{k}$ is assumed to be between $0.8$ and $1.5$. 
{Each element in the PCQ matrix is generated according to 
\eqref{eq:PCQ_indicator} with the input of $\alpha_{k}$, $\sigma_{s,k,f_n}^2$ and $f_n$.}

\subsection{Network Performance with Null-ACI system}
We first investigate the performance of the proposed Hungarian-based algorithm in Null-ACI system. We consider the following two benchmark algorithms for comparison purposes: 
\begin{itemize}
	\item \emph{Greedy Selection Scheme}: The greedy selection (GS) scheme is a heuristics algorithm, where the UAV is associated with the communication channel at each time slot according to the maximum PCQ criterion, i.e., choosing the minimum element in $\mathbf{\Phi}_s$. 
	Repeat the above procedure until all UAVs in the swarm are associated with a specific communication channel. The detail procedure is summarized in Algorithm \ref{algGS}.	
		
	\item \emph{Baseline Scheme 1}: We remark that Baseline scheme 1 is different from the GS scheme in that each UAV is randomly associated with the single communication channel at each time slot \cite{rand_base}. Note that in each time slot, each UAV only occupies one channel whilst each channel is only assigned to at most one UAV .
\end{itemize}

%a heuristic algorithm to problem \eqref{eq:MMF_NOACI4} based on greedy selection (GS). Specifically, this algorithm greedily chooses the UAV and channel corresponding to the minimum value in Table \ref{eq:assignment_table}, and continues to select among the remaining UAVs and channels until all of them are allocated. The detail procedure is presented in Algorithm \ref{algGS}.

\begin{algorithm}[!t]
	\caption{Proposed Greedy Selection Algorithm for Null-ACI system}
	\label{algGS}
	\begin{algorithmic}[1]
		\STATE \textbf{Initialize} $t_{2} = 1, \xi_{s} = 0$.
		\FOR{$t_{2} = 1$ to $K$}
		\STATE Select $(n^{*},k^{*}) = \mathop{\arg\min}\limits_{n\in\mathcal{N},k\in\mathcal{K}} \phi^n_{s,k}$ in the PCQ matrix $\mathbf{\Phi_{s}}$.\
		\STATE Record the optimal selection strategy: $(n^{*},k^{*})$.
		\STATE Compute  $ \xi_{s} =  \xi_{s} + \mathop{\min}\limits_{n\in\mathcal{N},k\in\mathcal{K}} \phi^n_{s,k}$.
		\STATE Update $\mathcal{N} \triangleq \mathcal{N} \setminus\lbrace n^{*} \rbrace$.\ 
		\STATE Update $\mathcal{K} \triangleq \mathcal{K} \setminus\lbrace k^{*} \rbrace$.\ 
		\ENDFOR
		\STATE Compute the optimal SINR value:	
		$\overline\gamma_{s} = \frac{P_{max}}{\xi_{s}}$.		
		\STATE Compute the optimal power allocation:
		\begin{center}
		$p_{s, k}^{*} = \overline\gamma_{s} \frac{ \sigma_{s, k,f_{n}^{*}}^2}{\alpha_k c_{s, k} (f_{n}^{*})^{-2}}$.
		\end{center}
	\end{algorithmic}
\end{algorithm}

\begin{figure}[!t]
	\centering
	\includegraphics[width=3.5in,height=3.0in]{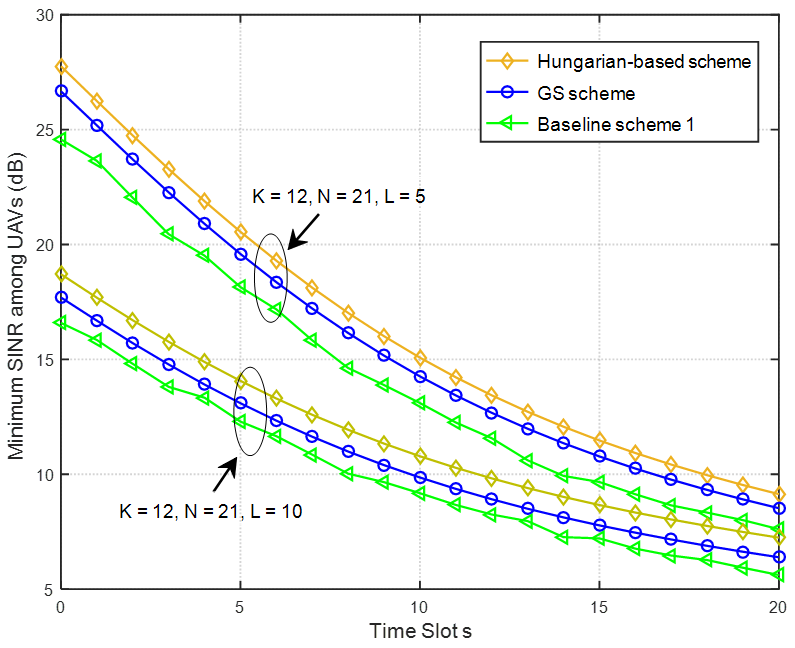}
	\caption{The minimum SINR among all the UAVs versus the different flight phases for different schemes in the Null-ACI system when $K=12$ and $N=21$.}\label{fig:NOACI_1}
\end{figure}

Fig. \ref{fig:NOACI_1} compares the communication performance of the UAV swarm over the different flight phases for the different schemes with different number of radiation sources $L$ when $K=12$ and $N=21$. 
It is observed that the communication quality of each UAV declines gradually as the flight time $s$ increases. This is because the UAV formation gradually moves away from the GCU after the mission starts, thereby resulting in a less favorable propagation condition over the time.
When the number of radiation sources increases, the intensity of EI accordingly enhances, and thus the minimum SINR among UAVs at each time slot achieved by all schemes gradually decreases.
In addition, we notice that the proposed Hungarian-based scheme outperforms all the other competing schemes in the entire flight duration. For instance, the proposed Hungarian-based scheme improves over the Baseline scheme 1, by around $23\%$ at time slot $20$. The reason for this outcome is that the proposed Hungarian-based scheme can exploit the distinguishing features
of different radiation sources and communication channels to effectively suppress the EI, while the other competing schemes do not take these features into consideration.

%swarm ($K = 12, N = 21$) for the proposed three algorithms, in the case of $L=5$ and $10$. 
%We first evaluate the performance trends during the flight. Apparently, as flight time increases, the UAV formation gradually moves away from the GBS. It is observed that the communication quality of UAVs declines gradually with the distance from the GBS increasing. 
%What's more, with the increasing number of radiation sources, the EI becomes stronger, so the SINR of UAV swarm decreases accordingly. 
%Morever, one can see that the performance gap between the three algorithms is not large, but the proposed Hungarian method always achieves the higher SINR, by up to approximately $123\%$ and $110\%$, compared to the GS algorithm and Baseline scheme 1. This improvement mainly comes from that the Hungarian algorithm makes the channel assignment scheme more reasonable, which is effective in alleviating EI in communication system. 

%One can see that the minimum SINR of the Hungarian algorithm is significantly higher than that of other two schemes: the improvement is approximately 21 percent and 10 percent when compared with GS method and Baseline scheme 1.
%As a result

\begin{figure}[!t]
	\centering
	\includegraphics[width=3.5in,height=3.0in]{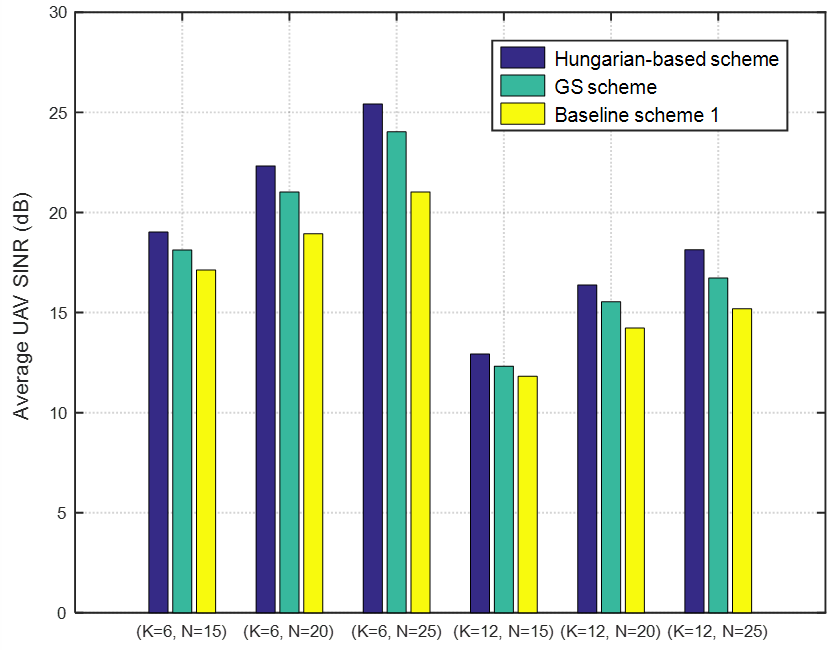}
	\caption{Average UAV SINR performance comparison under different scenarios in the Null-ACI system.}\label{fig:NOACI_2}
\end{figure}

In Fig. \ref{fig:NOACI_2}, we show the average UAV SINR performance comparison for different schemes under various system parameters (i.e., the number of UAVs $K$ and the number of channels $N$). For the sake of fairness, we assume that there are $L=5$ radiation sources distributed across the region with fixed locations. It shows that as more communication channels become available for a fixed number of UAVs, all the schemes considered can achieve better average UAV SINR performance. On the contrary, the average UAV SINR achieved by all the schemes will gradually deteriorate with the increasing number of UAVs in the formation when the system spectrum resource is limited.
In addition, we can observe that the proposed Hungarian-based algorithm achieves a significant gain over the competing schemes, which demonstrates the importance of the powerful channel assignment strategy. Moreover, as the number of communication channels increases, the performance gap between the proposed Hungarian-based scheme and the competing schemes becomes larger. Hence, it appears that for multi-UAV network with more available spectrum resources, our proposed Hungarian-based scheme is particular appealing from an optimum resource allocation perspective.

%offers small performance advantage over the other algorithms, and slowly increases with more bands available. For example, for $K=12$ and $N=15$, the SINR gaps between the Hungarian algorithm and other two algorithms are $5.2\%$ and $9.5\%$, respectively, reaching up to $7.3\%$ and $19.4\%$ in $K=12$ and $N=25$ system. The reason lies in that the Hungarian algorithm can obtain the global optimal solution, which offers superior performance as more communication resources are available.

% As a consequence
\begin{figure}[!t]
	\centering
	\includegraphics[width=3.5in,height=3.0in]{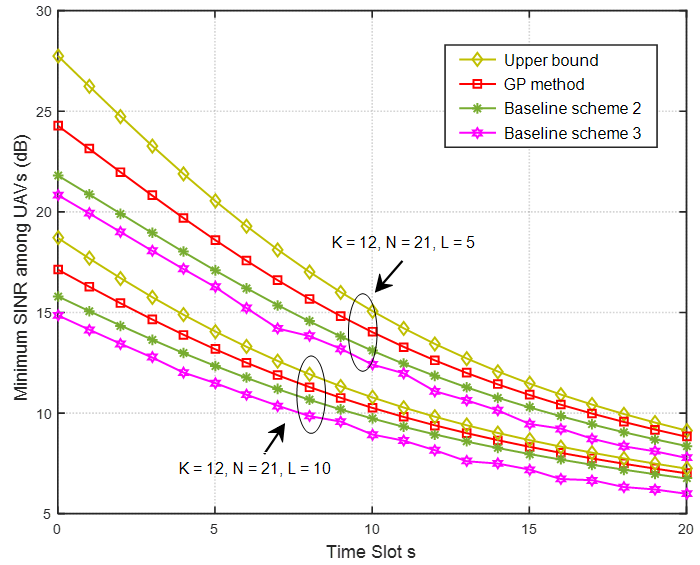}
	\caption{The minimum SINR among all the UAVs versus the different flight phases for different schemes in the ACI system when $K=12$ and $N=21$.}\label{fig:ACI_1}
\end{figure}

\begin{figure}[!t]
	\centering
	\includegraphics[width=3.5in,height=3.0in]{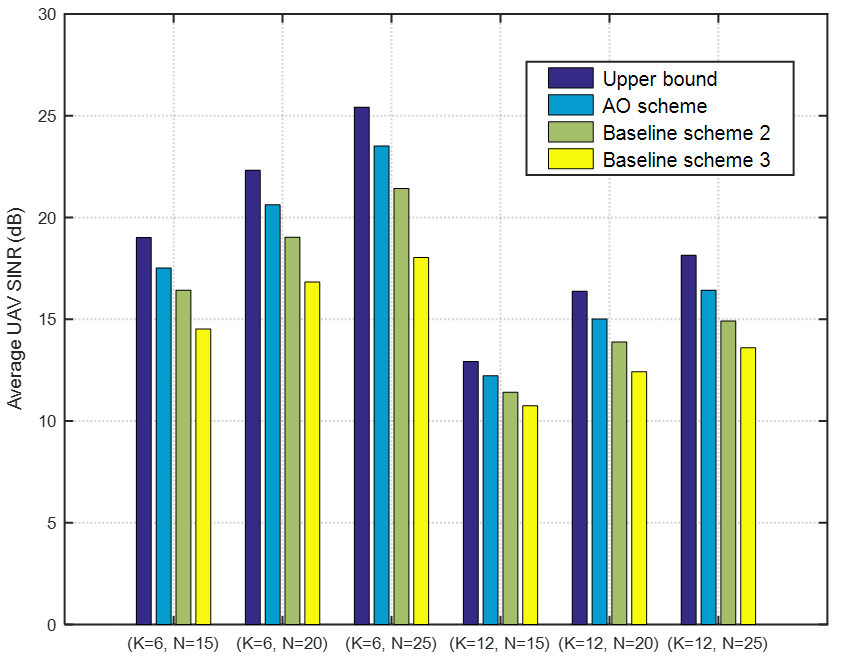}
	\caption{Average UAV SINR performance comparison under different scenarios in the ACI system.}\label{fig:ACI_2}
\end{figure}

\subsection{Network Performance with ACI system}
Now, we report simulation experiments to evaluate the performance of the proposed AO algorithm in ACI system. The following three resource allocation schemes are also simulated as benchmark for comparison purpose:
\begin{itemize}
	\item \emph{Upper Bound}: In this scheme, the ACI is assumed to be null, and we adopt the proposed Hungarian-based algorithm to obtain the optimal resource allocation strategy. Moreover, the performance of this scheme can be served as the performance upper bound for the ACI system.
%	We take the Hungarian-based scheme in the null-ACI system as a Upper bound for the ACI system. 	
%	The Null-ACI system can also be regarded as a special case in ACI communication system, in which the ACI is so weak that it closes to 0. As the Hungarian algorithm can get the global optimal solution of Null-ACI system, we take the performance of Hungarian algorithm without ACI case as the upper bound in ACI system.		
	\item \emph{Baseline Scheme 2}: 
	This scheme consists of two steps. Specifically, the first step is to use the same channel assignment method as in the upper bound scheme, while in the second step, the uplink transmission power of GCU is determined by the eigenvalue decomposition \cite{ZLY}.
%	Inspired by the former strategies, we introduce a simple algorithm (named Baseline scheme 2) in ACI system. This method consists of two steps, the first step is to adopt the same channel assignment method as the upper bound method, while in the second step, we optimize the uplink transmission power by the eigenvalue decomposition technique.
	\item \emph{Baseline Scheme 3}: 
	In this scheme, we randomly assign a communication channel to each UAV in sequence \cite{rand_base}. Based on the channel assignment results, the uplink transmission power of GCU is optimized through the eignevalue decomposition. 
%	To demonstrate the efficacy of our proposed channel assignment algorithm, we introduce a random algorithm (named Baseline scheme 3), where we perform a random channel assignment followed by the optimum power allocation through eigenvalue decomposition.
\end{itemize}

Fig. \ref{fig:ACI_1} plots the communication performance of the UAV swarm versus the different flight phases for different schemes with different number of radiation sources $L$ in the ACI system when $K=12$ and $N=21$. 
%Note that as the flight time $s$ elapses, the communication performance deteriorates gradually. This performance decline is due to the fact that 
%as the distance between the GBS and UAV increases, which will lead to the 
%and the network performance deteriorates gradually. This performance decline is due to the fact that 
%thanks to the control signals weakens and interference enhances. Specifically, 
Similar to Null-ACI network, it is observed that as UAV swarm moves away from the GCU, the network performance deteriorates gradually, owing to the increased control signal attenuation. Besides, we notice that there is a huge performance gap between the proposed AO scheme and the upper bound at the beginning of carrying out the mission. This is intuitive since the intensity of the ACI is proportional to the distance between the GCU and two UAVs in the adjacent channels, and its initial value is particular high. This fact inevitably leads to the performance degradation even with the aid of some interference management techniques.
Furthermore, it can be seen that the minimum SINR performance achieved by the proposed AO scheme is superior to that of baseline 2 and 3. This is because the proposed AO scheme can make full use of the entire network radio resources by leveraging the joint optimization of power control and channel assignment, and further achieves  more preferable fairness among the UAVs in a swarm.
%the control signals weakens and interference enhances. 
%In fact, the results in Fig. \ref{fig:ACI_1} show that the proposed AO scheme exploits the network resources more efficiently, compared to Baseline scheme 2 and 3. This performance gain is due to the fact that the proposed AO scheme can reasonably assign channels according to the intensity of ACI and EI, which can effectively combat interference and improve the signal quality.
%What's more, as the flight time $s$ increases, the Upper bound is gradually approached by GP scheme, which indicates that our proposed iterative algorithm achieves an ideal performance in communication network. 
%This is because the gradual increase in the distance between the GBS and UAV swarm will lead to the weakening of the ACI in the system, 
Fig. \ref{fig:ACI_2} intuitively shows the average UAV SINR performance for various schemes under different system parameters (i.e., the number of UAVs $K$ and the number of channels $N$), with $L = 5$ radiation sources randomly distributed at the predefined 2D region. 
It is interesting to note that the proposed AO scheme achieve higher minimum SINR than that of baseline 2 and 3, and the performance gain become more substantial when the number of UAVs served decreases or the number of available communication channels increases.

\begin{figure}[!t]
	\centering
	\includegraphics[width=3.5in,height=3.0in]{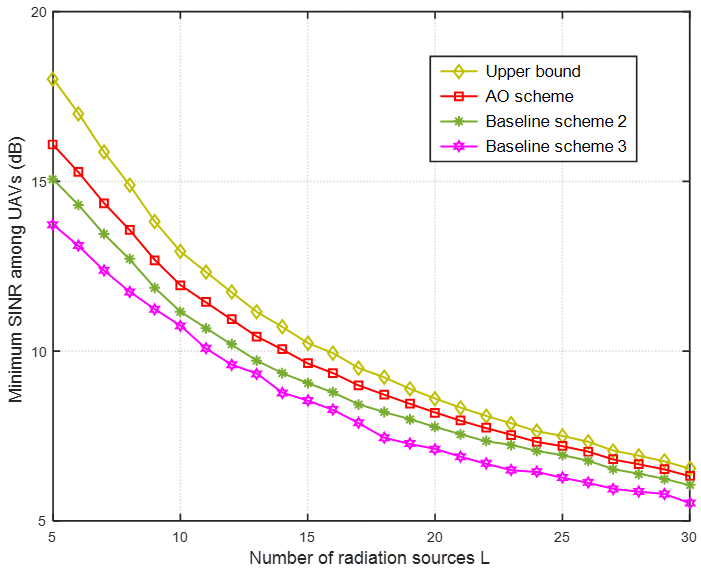}
	\caption{The minimum SINR among all the UAVs versus the number of radiation sources $L$ for different schemes in the ACI system when $K=12$ and $N=21$.
	}\label{fig:UAV_radiation}
\end{figure}

In Fig. \ref{fig:UAV_radiation}, we compare the minimum SINR among all the UAVs versus the number of radiation sources $L$ for different schemes when $K=12$ and $N=21$. We observe that the best minimum SINR performance among all the UAVs in a swarm is achieved by the upper bound scheme, followed by the proposed AO scheme. In addition, as more radiation sources are randomly generated in the network, the gap between the proposed AO scheme and the performance upper bound vanishes. This is because besides the EI arising from the radiation sources, the ACI also significantly affects the quality of communication links between UAVs and the GCU in the small and moderate $L$ regime.
Furthermore, the performance of baselines 2 and 3  is inferior to that of the proposed AO scheme due to the ineffective channel assignment strategies.
Last, the proposed AO scheme is observed to gradually approach the performance upper bound when $L$ is large, which implies that the proposed scheme can achieve a near-optimal performance in the interference-infested radio environment.

\begin{figure}[!t]
	\centering
	\includegraphics[width=3.5in,height=3.0in]{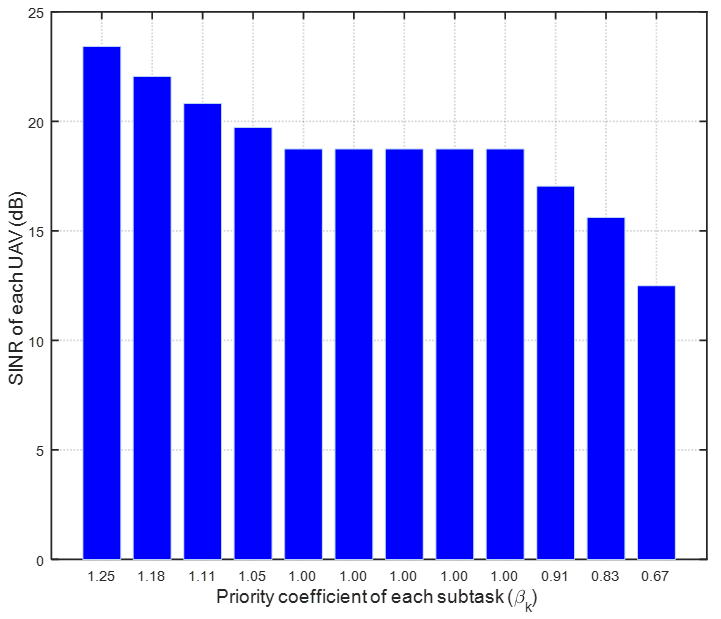}
	\caption{The impact of priority coefficients on the SINR performance of each UAV.}\label{fig:UAV_priority}
\end{figure}

Finally, we turn to investigate the impact of the sub-task weight factor on the SINR performance of each UAV. Note that different from the conventional weight factor in most of the existing works, in this paper a smaller value of $\alpha_k$ stands for a higher priority of the corresponding sub-task instead. To avoid any potential ambiguity, we hereby let $\beta_k\triangleq \alpha^{-1}_k$ be the priority coefficient of the subtask performed by the $k$-th UAV. Fig. \ref{fig:UAV_priority} depicts the SINR performance of each UAV versus the associated priority coefficient $\beta_k$ for the proposed AO algorithm with $K=12$, $N=25$, and $L=5$. We can observe that the SINR of each UAV is proportional to the value of priority coefficient $\beta_k$, which indicates that the UAV performing a more urgent subtask would be allotted with more favorable radio resources in the proposed AO algorithm. These results also demonstrate the efficiency of the proposed AO algorithm in handling different scenarios for task urgency (i.e., priority coefficient $\beta_k$) and its ability to strike a better balance among the UAVs in a swarm, thereby endowing added flexibility to the multi-UAV network.

%the priority on the performance of each UAV. It is worth emphasizing that the priority factor $\alpha_{k}$ in this paper is different with the traditional priority coefficient. Specifically, a low $\alpha_{k}$ represents a higher priority coefficient of UAV, hence we define the priority coefficient for UAV $k$ as $\alpha_{k}^{-1}$.
%We randomly select one time slot of the GP algorithm, at the scenario of $K=12, N=25$ and $L=5$.
%From the results of Fig. \ref{fig:UAV_priority}, we can observe that smaller $\alpha_{k}$ often leads to better performance. The main reason is as follows. Note that at the optimality of problem \eqref{eq:MMF-Ps} the component functions of the objective must achieve the same value \cite{Optimal_downlink_eig}. As a result, considering the $\alpha_{k}$ in the component functions, it is inversely proportional to the SINR of UAV.
%Therefore, in practical application, we can set smaller $\alpha_{k}$ for the host UAV, and larger $\alpha_{k}$ for the slave UAV, in order to preferentially maintain the performance of the host UAV.

\section{CONCLUSION}
This paper studied the performance of a GCU-to-UAV uplink communication system. To support reliable communication while effectively reducing the impact of ACI and EI, we proposed a priority-based resource coordination scheme, where the channel assignment and power allocation are jointly optimized to maximize the minimum SINR among multiple UAVs. According to the intensity of ACI, we consider the corresponding problem in two scenarios, i.e., Null-ACI and ACI systems. By exploring the particular problem structure in Null-ACI case, we recast the formulation into an equivalent yet more tractable assignment problem and obtain the global optimal solution via Hungarian algorithm, which reveals the performance upper bound of communication system. For general ACI systems, we proposed an efficient iterative algorithm for its solution based on smooth approximation and alternating optimization methods. Extensive simulation results demonstrate that the proposed algorithms can significantly enhance the minimum SINR among all the UAVs as compared to the existing solutions and adapt the allocation of communication resources to diverse mission priority. 
This paper aimed to shed more light on the design and performance analysis of the multi-UAV communication system, which can be extended in several interesting directions for the future work, including intelligent trajectory optimization, adaptive resource coordination, as well as advanced priority-aware design, for both uplink and downlink transmissions.

\end{document}